\documentclass[11pt, twoside, letter]{article}
\usepackage{amsthm, amsmath, amssymb, amsfonts, graphicx, epsfig}
\usepackage{accents}

\usepackage{bbm}
\usepackage{mathrsfs}

\usepackage{epstopdf}

\usepackage{graphicx}
\usepackage[font=sl,labelfont=bf]{caption}
\usepackage{subcaption}

\usepackage{float}
\restylefloat{table}

\usepackage{enumerate}

\usepackage[usenames,dvipsnames]{color}

\usepackage[colorlinks=true, pdfstartview=FitV, linkcolor=blue,
            citecolor=blue, urlcolor=blue]{hyperref}

\usepackage{url}
\makeatletter\def\url@leostyle{%
 \@ifundefined{selectfont}{\def\UrlFont{\sf}}{\def\UrlFont{\scriptsize\ttfamily}}} \makeatother

\usepackage[margin=1.0in, letterpaper]{geometry}

\newtheorem{theorem}{Theorem}
\newtheorem{proposition}[theorem]{Proposition}

\newtheorem{corollary}[theorem]{Corollary}

\theoremstyle{definition}
\newtheorem{definition}[theorem]{Definition}
\newtheorem{example}[theorem]{Example}
\theoremstyle{remark}
\newtheorem{remark}[theorem]{Remark}

\numberwithin{equation}{section}
\numberwithin{theorem}{section}

\definecolor{Red}{rgb}{0.9,0,0.0}
\definecolor{Blue}{rgb}{0,0.0,1.0}

\def\cA{\mathcal{A}}
\def\cB{\mathcal{B}}

\def\cD{\mathcal{D}}

\def\cF{\mathcal{F}}
\def\cG{\mathcal{G}}

\def\cM{\mathcal{M}}
\def\cN{\mathcal{N}}

\def\cP{\mathcal{P}}
\def\cQ{\mathcal{Q}}
\def\cR{\mathcal{R}}

\def\cT{\mathcal{T}}

\def\cX{\mathcal{X}}
\def\cY{\mathcal{Y}}

\def\bF{\mathbb{F}}

\def\bN{\mathbb{N}}

\def\bR{\mathbb{R}}

\def\bT{\mathbb{T}}

\def\bV{\mathbb{V}}

\def\bX{\mathbb{X}}

\newcommand{\1}{\mathbbm{1}}            
\newcommand{\set}[1]{\{#1\}}            
\newcommand{\Set}[1]{\left\{#1\right\}} 
\renewcommand{\mid}{\;|\;}              

\DeclareMathOperator*{\esssup}{ess\,sup} 
\DeclareMathOperator*{\essinf}{ess\,inf} 

\DeclareMathOperator{\Essinf}{ess\,inf} 
\DeclareMathOperator{\Esssup}{ess\,sup} 

\DeclareMathOperator{\var}{\mathrm{V}@\mathrm{R}}           
\DeclareMathOperator{\tvar}{\mathrm{TV}@\mathrm{R}}         

\usepackage{tikz}
\usepackage{verbatim}
\usetikzlibrary{shadows,arrows}
\pgfdeclarelayer{background}
\pgfdeclarelayer{foreground}
\pgfsetlayers{background,main,foreground}
\usetikzlibrary{positioning}
\usetikzlibrary{calc}
\usetikzlibrary{decorations.markings}

\tikzstyle{materia}=[draw, fill=blue!20, text width=8.0em, text centered,
  minimum height=2em,drop shadow]

\tikzstyle{practica} = [materia, text width=15em, minimum width=12em,
  minimum height=3em, rounded corners, drop shadow]
  \tikzstyle{practica0} = [materia,text width=7em, minimum width=7em,
  minimum height=3em, rounded corners,  fill=blue!5,]
  \tikzstyle{practica1} = [materia, text width=17em, minimum width=12em,
  minimum height=3em, rounded corners, drop shadow]

\tikzstyle{texto} = [above, text width=6em, text centered]
\tikzstyle{linepart} = [draw, thick, color=black!50, -latex', dashed]
\tikzstyle{line} = [draw, thick, color=black!50, -latex']
\tikzstyle{ur}=[draw, text centered, minimum height=0.01em]

\newcommand*{\StrikeThruDistance}{0.15cm}%
\tikzset{strike-arrow/.style={
    decoration={markings, mark=at position 0.5 with {
        \draw [thin,-]
            ++ (-\StrikeThruDistance,-\StrikeThruDistance)
            -- ( \StrikeThruDistance, \StrikeThruDistance);}
    },
    postaction={decorate},
}}

\usepackage{multirow}
\usepackage{titling}
\newcommand*\circled[1]{\tikz[baseline=(char.base)]{
    \node[shape=circle,draw,inner sep=.15em] (char) {#1};}}
\newcommand*\rectangled[1]{%
  \tikz[baseline=(R.base)]\node[draw,rectangle,inner sep=.2em](R) {#1};\!
}

\title{A survey of time consistency of dynamic risk measures and dynamic performance measures in discrete time: LM-measure perspective}

\makeatletter
\def\and{%
  \end{tabular}%
  \begin{tabular}[t]{c}}%
\def\@fnsymbol#1{\ensuremath{\ifcase#1\or a\or b\or c\or
   d\or e\or f\or g\or h\or i\else\@ctrerr\fi}}
\makeatother

\author{
        Tomasz R. Bielecki\,\thanks{Department of Applied Mathematics, Illinois Institute of Technology
       \newline \hspace*{1.45em}  10 W 32nd Str, Building E1, Room 208, Chicago, IL 60616, USA
       \newline \hspace*{1.45em}  Emails: \url{tbielecki@iit.edu} (T.R. Bielecki) and \url{cialenco@iit.edu} (I. Cialenco)
       \newline \hspace*{1.45em}  URLs: \url{http://math.iit.edu/\~bielecki}  and \url{http://math.iit.edu/\~igor}
        \vspace{0.5em}} ,
\and
        Igor Cialenco\,\footnotemark[1] ,
\and and
        Marcin Pitera\,\thanks{Institute of Mathematics, Jagiellonian University,  Cracow, Poland
        \newline \hspace*{1.45em}    Email: \url{marcin.pitera@im.uj.edu.pl}, URL: \url{http://www2.im.uj.edu.pl/MarcinPitera/} }
        }

\date{}
\date{ {\small
Forthcoming in \textit{Probability, Uncertainty and Quantitative Risk} \\[1em]
 First Circulated: March 29, 2016\\
 This Version: December 19, 2016
}}

\begin{document}

\maketitle


{\footnotesize
\begin{tabular}{l@{} p{350pt}}
  \hline \\[-.2em]
  \textsc{Abstract}: \ & In this work we give a comprehensive overview of the time consistency property of dynamic risk and performance measures, focusing on a the discrete time setup. The  two key operational concepts used throughout are the notion of the LM-measure and the notion of the update rule that, we believe,  are the key tools for studying time consistency in a unified framework. \\[0.5em]
\textsc{Keywords:} \ &  time consistency, update rule, dynamic LM-measure, dynamic risk measure, dynamic acceptability index, measure of performance. \\
\textsc{MSC2010:} \ & 91B30, 62P05, 97M30, 91B06. \\[1em]
  \hline
\end{tabular}
}

\begin{quote}
\textit{
``The dynamic consistency axiom turns out to be the heart of the matter.''}

\hfill A. Jobert and  L. C. G. Rogers

\hfill\tiny{Valuations and dynamic convex risk measures, Math Fin 18(1), 2008, 1-22.}
\end{quote}

\section{Introduction}\label{sec:Intro}

The goal of this work is to give a comprehensive overview of the time consistency property of dynamic risk and performance measures. We focus on discrete time setup, since most of the existing literature on this topic is dedicated to this case.

The  time consistency surveyed in this paper is related to dynamic decision making subject to various uncertainties that evolve in time. Typically, decisions are made subject to the decision maker's preferences, which may change in time and thus they need to be progressively assessed as an integral part of the decision making process. Naturally, the assessment of preferences should be done in such a way that the future preferences are assessed consistently with the present ones. This survey is focusing on this aspect of time consistency of a dynamic decision making process.

Traditionally, in finance and economics, the preferences are aimed at ordering cash and/or consumption streams. A convenient way to study preferences is to study them via numerical representations, such as (dynamic) risk measures, (dynamic) performance measures,  or, more generally, dynamic LM-measures\footnote{An LM-measure is a function that is \textit{local} and \textit{monotone}; see Definition~\ref{def:LM-measure}. These two properties  have to be satisfied by any reasonable dynamic measure of performance and/or measure of risk, and  are shared by most such measures in the existing literature.}  \cite{BCP2014a}. Consequently, the study of the time consistency of preferences is also conveniently done in terms of their numerical representations. This work is meant to survey various approaches to modelling and analysis of the time consistency of numerical representations of preferences.

As stated above, the objects of our survey---the dynamic LM-measures---are meant to ``put a preference order'' on the sets of underlying entities. There exists a vast literature on the subject of preference ordering, with various approaches towards establishing an order of choices, such as  the decision theory or the expected utility theory, that trace their origins to the mid 20th century.
We focus our attention, essentially, on the axiomatic approach to defining risk or performance measures.

The axiomatic approach to measuring risk of a financial position was initiated in the seminal paper by Artzner et al. \cite{ArtznerDelbaenEberHeath1999}, and has been going through a flourishing development since then.
The measures of risk introduced in \cite{ArtznerDelbaenEberHeath1999}, called coherent risk measures, were meant to determine the regulatory capital requirement by providing a numerical representation of the riskiness of a portfolio of financial assets.
In this framework, from mathematical point of view, the financial positions are understood as either discounted terminal values (payoffs) of  portfolios, that are modeled in terms of random variables, or they are understood as discounted dividend processes, cumulative or bullet, that are modeled as stochastic processes.
Although stochastic processes can be viewed as random variables (on appropriate spaces), and vice versa - random variables can be treated as particular cases of processes---it is convenient, and in some instances necessary, to treat these two cases separately---the road we are taking in this paper.

In the paper \cite{ArtznerDelbaenEberHeath1999}, the authors considered the case of random variables, and the risk measurement was done at time zero only. This amounts to considering a one period time model in the sense that the measurement is done today of the cash flow that is paid at some fixed future time (tomorrow). Accordingly, the related risk measures are referred to as static measures.
Since then, two natural paths were followed: generalizing the notion of risk measure by relaxing or changing the set of axioms, or/and considering a dynamic setup. By dynamic setup we mean that the measurements are done throughout time and are adapted to the flow of available information. In the dynamic setup, both  discrete and continuous time evolutions have been studied, for both  random variables and stochastic processes as the inputs. In the present work, we focus our attention on the discrete time setup, although we briefly review the literature devoted to continuous time.

This survey is organized as follows.
We start with the literature review relevant to the dynamic risk and performance measures focusing on the time consistency property in the discrete time setup. In Section~\ref{sec:prelims}, we set the mathematical scene; in particular, we introduce the main notations used in this paper and the notion of LM-measures. Section~\ref{sec:TimeCons} is devoted to the time consistency property. There we discuss  two generic approaches to time consistent assessment of preferences and point out several idiosyncratic approaches. We put forth in this section the notion of an update rule that, we believe,  is the key tool for studying time consistency in a unified framework.
Sections~\ref{s:tc.rv} and \ref{s:tc.sp} survey some concepts and results regarding time consistency in the case of random variables and in the case of stochastic processes, respectively. Our survey is illustrated by numerous examples that are presented in Section \ref{sec:Examples}. We end the survey with two appendices. In Appendix~\ref{S:A}  we provide a brief exposition of the three fundamental concepts used in the paper: the dynamic LM-measures, the conditional essential suprema/infima, and LM-extensions. Finally, in Appendix~\ref{A:proofs} we collect proofs of several results stated throughout our survey.

\section{Literature review}\label{s:litRev}
The aim of this section is to give a chronological survey of the developments of the theory of dynamic risk and performance measures. Although it is not an obvious task to establish the exact lineup, we tried our best to account for the most relevant works according to adequate chronological order.

We trace back the origins of the research regarding time consistency to Koopmans~\cite{Koopmans1960} who put on the precise mathematical footing, in terms of the utility function, the notion of persistency over time of the structure of preferences.

Subsequently, in the seminal paper, Kreps and Porteus~\cite{KrepsPorteus1978} treat the time consistency at a general level by axiomatising the ``choice behavior'' of an agent by taking into account how choices at different times are related to each other; in the same work, the authors discuss the motivations for studying the dynamic aspect of choice theory.

Before we move on to reviewing the works on dynamic risk and performance measures, it is worth mentioning  that the robust expected utility theory proposed by Gilboa and Schmeidler~\cite{GilboaSchmeidler1989} can be viewed as a more comprehensive theory than the one discussed in \cite{ArtznerDelbaenEberHeath1999}; we refer to \cite{RoordaSchumacherEngwerda2005} for the relevant discussion.

Starting with \cite{ArtznerDelbaenEberHeath1999}, the axiomatic theory of risk measures, understood as functions mapping random variables into real numbers, was developing around the following main goals: a) to define a set of properties (or axioms) that a risk measure should satisfy; b) to characterize all functions that satisfy these properties; c) provide particular examples of such functions. Each of the imposed axioms should have a meaningful financial or actuarial interpretation. For example, in \cite{ArtznerDelbaenEberHeath1999}, a static coherent risk measure is defined as a function $\rho:L^\infty\to[-\infty, \infty]$ that is monotone decreasing (larger losses imply larger risk), cash-additive (the risk is reduced by the amount of cash added to the portfolio today), sub-additive (a diversified portfolio has a smaller risk) and positive homogenous (the risk of a rescaled portfolio is rescaled correspondingly), where $L^\infty$ is the space of (essentially) bounded random variables on some probability space $(\Omega, \cF,P)$\footnote{In the original paper \cite{ArtznerDelbaenEberHeath1999},  the authors considered finite probability spaces, but later the theory was elevated to a general probability space \cite{Delbaen2000,Delbaen2002}.}.
The descriptions or the representations of these functions, also called robust representations, usually are derived via duality theory in convex analysis, and are necessary and sufficient in their nature. Traditionally,  among such representations we find: representations in terms of the level or the acceptance sets; numerical representations in terms of the dual pairings (e.g., expectations).
For example, the coherent risk measure $\rho$ mentioned above can be described  in terms of its acceptance set
$\cA_\rho=\set{X\in L^\infty \mid \rho(X)\leq 0}$. As it turns out, the acceptance set $\cA_\rho$ satisfies certain characteristic properties, and any set $\cA$ with these properties generates a coherent risk measure via the representation $\rho(X)= \inf \set{m\in \bR \mid m +X\in\cA}$.  Alternatively,  the function $\rho$ is a coherent risk measure if and only if there exists a nonempty set $\cQ$ of probability measures, absolutely continuous with respect to $P$, such that
\begin{equation}\label{eq:crmStatic}
  \rho(X) = -\inf_{Q\in\cQ}E^{Q}[X].
\end{equation}
 The set $\cQ$ can be viewed as a set of generalized scenarios, and a coherent risk measure is equal to the worst expected loss under various scenarios. By relaxing the set of axioms, the static coherent risk measures were generalized to static convex risk measures and to an even more general class called monetary risk measures. See, for instance, \cite{Szego2002} for a survey of static risk measures, as well as \cite{CheriditoLi2009,CheriditoLi2008}. On the other hand, axiomatic theory of performance measures was originated in \cite{ChernyMadan2009}.  A general  theory of risk preferences and their robust representations, based on only two generic axioms, was studied in \cite{Drapeau2010,DrapeauKupper2010}.

Moving to the dynamic setup, we first introduce an underlying  filtered probability space \linebreak $(\Omega,\cF,\set{\cF_t}_{t\geq 0}, P)$, where the increasing  collection of $\sigma$-algebras $\cF_t, \ t\geq 0,$ models the flow of information that is accumulated through time.

Artzner et al. \cite{ArtznerDelbaenEberHeathKu2002a} and \cite{ArtznerDelbaenEberHeathKu2002b} study an extension of the static models examined in  \cite{ArtznerDelbaenEberHeath1999} to the multiperiod case, assuming discrete time and discrete probability space.
The authors proposed a method of constructing dynamic risk measures $\set{\rho_t: L^\infty(\cF_T)\to \bar L^0(\cF_t), \ t=0,1\ldots,T}$,  by a backward  recursion, starting with $\rho_T(X)=-X$, and letting
\begin{equation}\label{eq:1}
\rho_t(X) = -\inf_{Q\in\cQ}E^{Q}[-\rho_{t+1} (X) \mid \cF_t], \quad 0\leq t< T,
\end{equation}
where, as before, $\cQ$ is a set of probability measures. If, additionally, $\cQ$ satisfies a  property called recursivity or consistency (cf. \cite{Riedel2004}), namely
\begin{equation}\label{eq:m-stable}
 \inf_{Q\in\cQ}E^Q[Z\mid \cF_t] = \inf_{Q\in\cQ} E^{Q}[\inf_{Q_1\in\cQ}E^{Q_1}[Z\mid \cF_{t+1}] \mid \cF_t], \quad t=0,1,\ldots, T-1, \ Z\in L^\infty,
\end{equation}
then one can show that \eqref{eq:1} is equivalent to
\begin{equation}\label{eq:2}
  \rho_{t}(X) = \rho_t(-\rho_{t+1}(X)), \quad 0\leq t <T, \ X\in L^\infty(\cF_T).
\end{equation}
The property \eqref{eq:2} represents what has become known in the literature as the \textit{strong time consistency} property.
{For example, if $\cQ=\set{P}$, then the strong time consistency reduces to the tower property for conditional expectations.}
From a practical point of view, this property essentially means that assessment of risks propagates in a consistent way over time: assessing at time $t$ future risk, represented by random variable $X$, is the same as assessing at time $t$ a risky assessment of $X$ done at time $t+1$ and represented by $-\rho_{t+1}(X).$
Additionally, the property \eqref{eq:2} is closely related to the Bellman principle of optimality or to the dynamic programming principle (see, for instance, \cite{BellmanDreyfus1962,Carpentier2012}).

Delbaen~\cite{Delbaen2006} studies the recursivity property in terms of $m$-stable sets of probability measures, and also describes the time consistency of dynamic coherent risk measures in the context of martingale theory.  The recursivity property is equivalent to properties known as time consistency and the rectangularity in the multi-prior Bayesian decision theory. Epstein and Schneider \cite{EpsteinSchneider2003} study time consistency and rectangularity  property  in the framework of ``decision under ambiguity.''

It needs to be said that several authors refer to  \cite{Wang1999} for an alternative axiomatic approach to time consistency of dynamic risk measures.

The first study of dynamic risk measures for stochastic processes (finite probability space and discrete time) is attributed to Riedel~\cite{Riedel2004}, where the author introduced  the (strong) time consistency as one of the axioms. If $\rho_t, \ t=0,\ldots,T,$ is a dynamic coherent risk measure, acting on the set of discounted terminal cash flows\footnote{In \cite{Riedel2004}, the author considered discounted dividend processes, but for simplicity here we write the time consistency for random variables.}, then $\rho$ is \textit{strongly time consistent}  if the following implication holds true:
\begin{equation}\label{eq:strongRiedel}
\rho_{t+1}(X)=\rho_{t+1}(Y) \ \Rightarrow \ \rho_t(X)=\rho_t(Y).
\end{equation}
This means that if tomorrow  we assess  the riskiness of $X$ and $Y$ at the same level, then today $X$ and $Y$ must have the same level of riskiness. It can be shown that for dynamic coherent risk measures, or more generally for dynamic monetary risk measures, property \eqref{eq:strongRiedel} is equivalent to \eqref{eq:2}.

Motivated by results regarding  the pricing procedure in incomplete markets, based on use of risk measures,  Roorda et al. \cite{RoordaSchumacherEngwerda2005} study dynamic coherent risk measures (for the  case of random variables on finite probability space and discrete time) and introduce the notion of (strong) time consistency; note  that their work was similar and contemporaneous to \cite{Riedel2004}. They show that  strong time consistency entails recursive computation of the corresponding optimal hedging strategies. Moreover, time consistency is also described in terms of the collection of probability measures that satisfy the ``product property,'' similar to the rectangularity property mentioned above.

Similarly, as in the static case, the dynamic coherent risk measures were extended to dynamic convex risk measures by replacing sub-additivity and positive homogeneity properties with convexity. In the continuous time setup, Rosazza Gianin \cite{RosazzaPhDThesis2002} links dynamic convex risk measures to nonlinear expectations or $g$-expectations, and to Backward Stochastic Differential Equations (BSDEs). Strong time consistency plays a crucial role and, in view of \eqref{eq:2}, it is equivalent to the tower property for conditional $g$-expectations.
These results are further studied in a sequel of papers \cite{RosazzaGianin2006,Peng2004,FrittelliGianin2004}, as well as in Coquet et al. \cite{CoquetHuMeminPeng2002}.

A representation similar to \eqref{eq:crmStatic} holds true for dynamic convex risk measure
 \begin{equation}\label{eq:dconvex}
  \rho_t(X) = - \inf_{Q\in\cM(P)} \left( E^Q[X\mid \cF_t] +\alpha_t^{\min}(Q)\right), \quad t=0,1,\ldots,T,
\end{equation}
where $\cM(P)$ is the set of all probability measures absolutely continuous with respect to $P$, and $\alpha^{\min}$ is the  minimal penalty function.\footnote{See Appendix~\ref{A:robust} for the definition of minimal penalty functions, up to a sign, and for the corresponding robust representations.} The natural question of describing (strong) time consistency in terms of properties of the minimal penalty functions was studied by Scandolo~\cite{Scandolo2003PhD}. Also in \cite{Scandolo2003PhD}, the author discusses the importance in the dynamic setup of the special property called locality. It should be mentioned that locality property was part of the earlier developments in the theory of dynamic risk measures. For example, it was called dynamic relevance axiom in \cite{Riedel2004}, and zero-one law in \cite{Peng2004}. Similarly to previous studies, \cite{Scandolo2003PhD} finds a relationship between time consistency, the recursive construction of dynamic risk measures, and the supermartingale property.  These results are further investigated in Detlefsen and Scandolo~\cite{DetlefsenScandolo2005}. Also in these works, it was shown that the dynamic entropic risk measure is a strongly time consistent convex risk measure.

Weber~\cite{Weber2006} continues the study of dynamic convex risk measures for random variables in a discrete time setup and introduces weaker notions of time consistency  \textit{acceptance and rejection time consistency}. Mainly, the author studies the law invariant risk measures, and  characterizes  time consistency in terms of the acceptance indicator $a_t(X) = \1_{\rho_t(X)\leq 0}$ and in terms of the acceptance sets of the form $\cN_t=\set{X \mid \rho_t(X)\leq 0}$.
Along the same lines, F\"ollmer and Penner~\cite{FollmerPenner2006} investigate the dynamic convex risk measures, representation of strong time consistency as a recursivity property, and they relate it to the Bellman principle of optimality. They also prove that
the supermartingale property of the penalty function corresponds to the weak or acceptance/rejection time consistency.
Moreover, the authors study  the co-cycle property of the penalty function for the dynamic convex risk measures that admit robust representation (see Definition~\ref{eq:representable}).

Artzner et al. \cite{ArtznerDelbaenEberHeathKu2007} continue to study the strong time consistency for dynamic risk measures, its equivalence with the stability property of test probabilities and with the optimality principle.

It is worth mentioning that Bion-Nadal~\cite{Bion-Nadal2004} studies dynamic monetary risk measures in a continuous time setting and their time consistency property in the context of  model uncertainty when the class of probability measures is not specified.

Motivated by optimization subject to risk criterion, Ruszczynski and Shapiro~\cite{RuszczynskiShapiro2006a} elevate the concepts from \cite{RuszczynskiShapiro2006} to the dynamic setting, with the main goal to establish conditions under which the dynamic programming principle holds.

Cheridito and Kupper~\cite{CheriditoKupper2006} introduce the notion of aggregators and generators for dynamic convex risk measures and give a thorough discussion about the composition of time-consistent convex risk measures in the discrete time setup, for both random variables and stochastic processes. They link time consistency to one step dynamic penalty functions. In this regard, we also refer to \cite{CheriditoDelbaenKupper2006,CheriditoKupper2009}.

Jobert and Rogers~\cite{JobertRogers2008} take the valuation concept as the starting point, rather than the dynamics of acceptance sets, with the valuation functional being the negative of a risk measure.  To quote the authors (strong) ``time consistency is the heart of the matter.'' Kloeppel and Schweizer~\cite{KloeppelSchweizer2007} use dynamic convex risk measures for valuation in incomplete markets, where the time consistency plays a key role. Cherny~\cite{Cherny2007} uses dynamic coherent risk measure for pricing and hedging European options; see also \cite{cherny-2006p5}.

Roorda and Schumacher~\cite{RoordaSchumacher2007} study the weak form of time consistency for dynamic convex risk measure.

Bion-Nadal~\cite{Bion-Nadal2006} continues to study various properties of dynamic risk measures, both in discrete and in continuous time, mainly focusing on the composition property mentioned above, and thus on the strong time consistency. The composition property is characterized in terms of stability of probability sets. The author defines the  \textit{co-cycle condition for the penalty function} and shows its equivalence to strong time consistency. In the followup paper, \cite{Bion-Nadal2008}, the author continues to study the characterization of time consistency in terms of the co-cycle condition for minimal penalty function.  For further related developments in the continuous time framework see \cite{BionNadal2009}.

Observing that Value at Risk ($\var$) is not strongly time consistent, Boda and Filar~\cite{BodaFilar2006}, and Cheridito and Stadje~\cite{CheriditoStadje2009} construct a strongly time consistent alternative to $\var$ by using a recursive composition procedure.

Tutsch~\cite{Tutsch2008} gives a different perspective on time consistency of convex risk measures by introducing the update rules\footnote{In the present manuscript, we also use the name `update rules', although the concept used here is different from that introduced in \cite{Tutsch2008}.} and generalizes the strong and weak form of time consistency via test sets.

The theory of dynamic risk measures finds its application in areas beyond the regulatory capital requirements. For example, Cherny~\cite{Cherny2010} applies dynamic coherent risk measures to risk-reward optimization problems and in \cite{Cherny2009a} to capital allocation;  Bion-Nadal~\cite{BionNadal2009a} uses dynamic risk measures for time consistent pricing;
Barrieu and El Karoui~\cite{Barrieu2004,Barrieu2005,Barrieu2007IP} study optimal derivatives design under dynamic risk measures;
Geman and Ohana~\cite{GemanOhana2008} explore the time consistency in managing a commodity portfolio via dynamic risk measures;
Zariphopoulou and Zitkovic~\cite{ZariphopoulouZitkovic2007} investigate the maturity independent dynamic convex risk measures.

In Delbaen et al. \cite{DelbaenPengGianin2010}, the authors establish a representation of the penalty function of dynamic convex risk measure using $g$-expectation  and its relation to the strong time consistency.

There exists a significant literature on a special class of risk measures that  satisfy the law invariance property.
Kupper and Schachermayer~\cite{KupperSchachermayer2009} prove that the only relevant, law invariant, strongly time consistent risk measure is the entropic risk measure.

For a fairly general study of dynamic convex  risk measures and their time consistency we refer to \cite{Bion-NadalKervarec2010} and \cite{Bion-NadalKervarec2010a}.
Acciaio et al. \cite{AcciaioFollmerPenner2010} give a comprehensive study of various forms of time consistency for dynamic convex risk measures in a discrete time setup. This includes  strong and weak time consistency, representations of time consistency in terms of acceptance sets, and the supermartingale property of the penalty function.
We would like to point out the  survey by Acciaio and Penner~\cite{AcciaioPenner2010} of discrete time dynamic convex risk measures. This work deals with (essentially bounded) random variables and examines most of the papers mentioned above from the perspective of the robust representation framework.

Although the connection between BSDEs and the dynamic convex risk measures in a continuous time setting had been established for some time, it appears that Stadje~\cite{Stadje2010} was the first author to create a theoretical framework for studying dynamic risk measures in discrete time via the Backward Stochastic Difference Equations (BS$\Delta$Es). Due to the backward nature of BS$\Delta$Es, the strong time consistency of risk measures played a critical role in characterizing the dynamic convex risk measures as solutions of BS$\Delta$Es. In a series of papers, Cohen and Elliott further studied the connection between dynamic risk measures and  BS$\Delta$Es \cite{CohenElliott2011,CohenElliott2009a,ElliottSiuCohen2011}.

F\"ollmer and Penner~\cite{FoellmerPenner2011} developed the theory of dynamic monetary risk measures under Knightian uncertainty, where the corresponding probability measures are not necessarily absolutely continuous with respect to the reference measure.
See also Nutz and Soner \cite{NutzSoner2010} for a study of dynamic risk measures under volatility uncertainty and their connection to $G$-expectations.

From a slightly different point of view, Ruszczynski~\cite{Ruszczynski2010} studies Markov risk measures, that enjoy strong time consistency, in the framework of risk-averse preferences; see also \cite{Shapiro2009,Shapiro2011,Shapiro2012,FanRuszczynski2014}.  Some concepts from the theory of dynamic risk measures are adopted to the study of the dynamic programming for Markov decision processes.

In the recent paper, Mastrogiacomo and Rosazza Gianin~\cite{MastrogiacomoRosazza2015} provide several forms of time consistency for sub-additive dynamic risk measures and their dual representations.

Finally, we want to mention that during the last decade significant advances were made towards developing a general theory of set-valued risk measures \cite{HamelRudloff2008,HamelHeydeRudloff2011,FeinsteinRudloff2013,HamelRudloffYankova2013,FeinsteinRudloff2015}, including the dynamic version of them, where mostly the corresponding form of strong time consistency is considered.

We recall that the main objective of use of risk measures for financial applications is mapping the level of risk of a financial position to a regulatory monetary amount expressed in units of the relevant currency. Accordingly, the key property of any risk measure is cash-additivity $\rho(X-m)=\rho(X)+m$. Clearly,  one can think of the risk measures as generalizations of $\var.$

A concept that is, in a sense, complementary to the concept of risk measures, is that of performance measures, which can be thought as generalizations of the well known Sharpe ratio.  In similarity with the theory of risk measures, the development of the theory of performance measures followed an axiomatic approach.  This was initiated by Cherny and Madan~\cite{ChernyMadan2009}, where the authors introduced the (static) notion of the coherent acceptability index--a function on $L^\infty$ with values in $\bR_+$ that is monotone, quasi-concave, and scale invariant. As a matter of fact, scale invariance is the key property of acceptability indices that distinguishes them from risk measures, and, typically, acceptability indices are not cash-additive.
The dynamic version of coherent acceptability indices was introduced by Bielecki et al.~\cite{BCZ2010}, for the case of stochastic processes, finite probability space, and discrete time. From now on, we will use as synonyms the terms \textit{measures of performance}, \textit{performance measures}, and \textit{acceptability indices}.

As it turns out, the time consistency for measures of performance is a delicate issue. None of the forms of time consistency, which had been coined for dynamic risk measures, are appropriate for dynamic performance  measures. In \cite{BCZ2010}, the authors introduce a new form of time consistency that is suitable for dynamic coherent acceptability indices. Let $\alpha_t, \ t=0,1,\ldots,T$, be a dynamic coherent acceptability index acting on $L^\infty$ (i.e., discounted terminal cash flows). We say that $\alpha$ is time consistent if the following implications hold true:
\begin{align}
 \alpha_{t+1}(X)\geq m_t &\quad \Rightarrow \quad \alpha_t(X)\geq m_t, \nonumber \\
 \alpha_{t+1}(X)\leq n_t &\quad \Rightarrow \quad \alpha_t(X)\leq n_t,  \label{eq:timeConstAlpha1}
\end{align}
where $X\in L^\infty$, and $m_t,n_t$ are $\cF_t$-measurable random variables.
Biagini and Bion-Nadal~\cite{BiaginiBion-Nadal2012} study dynamic  performance measures in a fairly general setup that generalize the results of \cite{BCZ2010}.
Later, using the theory of dynamic coherent acceptability indices developed in \cite{BCZ2010}, Bielecki et al.~\cite{BCIR2012} propose a pricing framework, called dynamic conic finance,  for dividend paying securities in discrete time. The time consistency property was at the core of establishing the connection between dynamic conic finance and classical arbitrage pricing theory. The static conic finance, that served as motivation  for \cite{BCIR2012}, was introduced in \cite{MadanCherny2010}. Finally, in recent papers \cite{BCC2014,RosazzaGianinSgarra2012}, the authors elevate the notion of dynamic coherent acceptability indices to the case of sub-scale invariant performance measures. For that, BSDEs are used in \cite{RosazzaGianinSgarra2012} and BS$\Delta$Es are used in \cite{BCC2014}.

For a general theory of robust representations of quasi-concave maps  that covers both dynamic risk measures and dynamic acceptability indices, see  \cite{FrittelliMaggis2011,BCDK2013,FrittelliMaggis2014,Bion-Nadal2016}.  Also in \cite{BCDK2013}, the authors study the strong time consistency of quasi-concave maps via the concept of certainty equivalence; see also \cite{FrittelliMaggis2010}.

 To our best knowledge, \cite{BCP2014a} is the only paper that combines into a unified framework the time consistency for dynamic risk measures  and dynamic performance measure. It uses the concept of update rules that serve as a vehicle  for connecting preferences at different times. We take the update rules perspective as the main tool for surveying the existing forms of time consistency.

We conclude this literature review by listing  works, which in our opinion,  are most relevant to this survey (not all of which are mentioned above though).

\smallskip

\noindent
\textbf{Dynamic Coherent Risk Measures}
\vspace{-.5em}
\begin{itemize}\itemsep-2pt
\item random variables, strong time consistency:
        \cite{ArtznerDelbaenEberHeathKu2002a}, \cite{ArtznerDelbaenEberHeathKu2002b},
        \cite{RoordaSchumacherEngwerda2005}.

\item stochastic processes,  strong time consistency:
        \cite{Riedel2004}, \cite{ArtznerDelbaenEberHeathKu2007}.

\end{itemize}

\noindent
\textbf{Dynamic Convex Risk Measures,}
\vspace{-.5em}
\begin{itemize}\itemsep-2pt
 \item random variables, strong time consistency:
(discrete time)  \cite{Scandolo2003PhD}, \cite{DetlefsenScandolo2005}, \cite{BodaFilar2006}, \cite{FrittelliScandolo2006},
                 \cite{RuszczynskiShapiro2006a},
                 \cite{FollmerPenner2006}, \cite{CheriditoStadje2009}, \cite{Bion-Nadal2006}, \cite{GemanOhana2008},  \cite{Bion-Nadal2008}, \cite{CheriditoKupper2009},
                 \cite{KupperSchachermayer2009}, \cite{AcciaioPenner2010}, \cite{Stadje2010},  \cite{AcciaioFollmerPenner2010},
                 \cite{CohenElliott2011}, \cite{CohenElliott2009a}, \cite{ElliottSiuCohen2011}
                 \cite{FasenSvejda2012}, \cite{BCDK2013}, \cite{BCP2014a}, \cite{IancuPetrikSubramanian2015}, \cite{MastrogiacomoRosazza2015}, \cite{RoordaSchumacher2015};         \\
                 (continuous time) \cite{RosazzaPhDThesis2002}, \cite{RosazzaGianin2006},
        \cite{FrittelliGianin2004}, \cite{Barrieu2004} \cite{DelbaenPengGianin2010}, \cite{KloeppelSchweizer2007},
        \cite{Bion-Nadal2006}, \cite{Barrieu2007IP}, \cite{Bion-Nadal2008}, \cite{Jiang2008}, \cite{Delbaen2012Book}, \cite{BionNadal2009}, \cite{Bion-NadalKervarec2010}, \cite{SircarSturm2015},  \cite{NutzSoner2010}, \cite{PennerReveillac2014}.

\item
random variables, supermartingale time consistency: \cite{Scandolo2003PhD}, \cite{DetlefsenScandolo2005}.
\item
random variables, acceptance/rejection time consistency:
    \cite{Weber2006}, \cite{FollmerPenner2006}, \cite{AcciaioFollmerPenner2010}, \cite{RoordaSchumacher2007}, \cite{Tutsch2008},
    \cite{AcciaioFollmerPenner2010}, \cite{BCP2014a}, \cite{RoordaSchumacher2015}.

\item
stochastic processes, strong and supermartingale time consistency: (discrete time) \cite{Scandolo2003PhD}, \cite{BCP2014a},
(continuous time) \cite{JobertRogers2008}

\end{itemize}

\noindent
\textbf{Dynamic Monetary Risk Measures,}
 strong time consistency:

 (discrete time)  \cite{CheriditoKupper2006}, \cite{CheriditoDelbaenKupper2006};
            (continuous time) \cite{Bion-Nadal2004}, \cite{FoellmerPenner2011}.

\smallskip
\noindent
\textbf{Dynamic Acceptability Indices:}
 \cite{BCZ2010}, \cite{BiaginiBion-Nadal2012},  \cite{BCIR2012}, \cite{RosazzaGianinSgarra2012}, \cite{BCDK2013}, \cite{FrittelliMaggis2014}, \cite{BCP2014a}, \cite{BCC2014}.

\section{Mathematical Preliminaries}\label{sec:prelims}
Let $(\Omega,\mathcal{F},\bF=\{\mathcal{F}_{t}\}_{t\in\mathbb{T}} ,P)$ be a filtered probability space, with $\mathcal{F}_{0}=\{\Omega,\emptyset\}$,  and $\bT=\set{0,1,\ldots, T}$, where $T\in\bN$ is a fixed and finite time horizon. We will also use the notation $\bT'=\set{0,1,\ldots, T-1}$.

For $\cG\subseteq\cF$ we denote by $L^0(\Omega,\cG,P)$ and $\bar{L}^0(\Omega,\cG,P)$ the sets of all $\cG$-measurable random variables with values in $(-\infty,\infty)$ and $[-\infty,\infty]$, respectively.
In addition, we use the notation $L^{p}(\cG):=L^{p}(\Omega,\cG,P)$, $L^{p}_{t}:=L^{p}(\mathcal{F}_{t})$, and $L^{p}:=L_T^{p}$, for $p\in \set{0,1,\infty}$; analogously we define $\bar{L}^{0}_t$.
 We also use the notation $\bV^{p}:=\{(V_{t})_{t\in\bT}: V_{t}\in L^{p}_{t}\}$, for $p\in \{0,1,\infty\}$.\footnote{Unless otherwise specified, it will be understood in the rest of the paper that $p\in \{0,1,\infty\}$. }  {Moreover, we use $\cM(P)$ to denote the set of all probability measures on $(\Omega,\cF)$ that are absolutely continuous with respect to $P$, and we set $\cM_t(P):=\{Q\in \cM(P)\,:\, Q|_{\cF_{t}}=P|_{\cF_{t}}\}$.}

Throughout this paper, $\cX$ relates to either the space of random variables $L^{p}$, or the space of adapted processes $\bV^p$.
If $\cX=L^{p}$,  then the  elements $X \in\cX$ are interpreted as discounted terminal cash flow. On the other hand, if $\cX=\bV^{p}$,  then the elements of $\cX$ are interpreted as discounted dividend processes.
All concepts developed for $\cX=\bV^{p}$ can be easily adapted to the case of the cumulative discounted value processes.
The case of random variables can be viewed as a particular case of stochastic processes by considering cash flow  with only the terminal payoff, i.e., stochastic processes such that $V=(0,\ldots,0,V_T)$. Nevertheless, we treat this case separately for transparency.
In both cases, we consider the standard pointwise order, understood in the almost sure sense.
In what follows, we also make use of the multiplication operator denoted as $\cdot_{t}$ and defined by:
\begin{align}
m\cdot_{t}V &:=(V_{0},\ldots,V_{t-1},mV_{t},mV_{t+1},\ldots), \nonumber\\
m\cdot_{t}X &:= mX,\label{eq:conventionV}
\end{align}
for  $V\in\Set{(V_t)_{t\in\bT} \mid V_{t}\in L^{0}_{t}}$, $X\in L^{0}$, $m\in L^{\infty}_{t}$, and $t\in\bT$.
In order to ease the notation, if no confusion arises, we drop $\cdot_t$ from the above product, and we simply write $mV$ and $mX$ instead of $m\cdot_{t}V$ and $m\cdot_{t}X$, respectively. For any $t\in\bT$ we set
\[
1_{\{t\}}:=
\begin{cases}
(\underbrace{0,0,\ldots,0}_{t},1,0,0,\ldots,0), & \textrm{if } \cX=\bV^p,  \\
1 & \textrm{if } \cX=L^p.
\end{cases}
\]
For any $m\in \bar L^{0}_t$, the value $m1_{\{t\}}$ corresponds to a cash flow of size $m$ received at time $t$. We use this notation for the case of random variables to present more unified definitions (see Appendix~\ref{S:families}).

\begin{remark}
We note that the space $\bV^{p}$,  endowed with  the multiplication $\,\cdot_{t}\,$, does not define a proper $L^{0}$--module \cite{FilipovicKupperVogelpoth2009,Vogelpoth2009PhD} (e.g., in general,  $0\cdot_{t} V\ne0$). {However, in what follows, we will adopt some concepts from $L^0$-module theory, which naturally fit into our study.
We refer the reader to \cite{BCDK2013,BCP2013} for a thorough discussion on this matter.}
\end{remark}

We use the convention  $\infty-\infty=-\infty+\infty=-\infty$ and $0\cdot\pm\infty=0$. Note that the distributive law does not hold true in general: $(-1)(\infty-\infty) = \infty \neq -\infty + \infty =-\infty. $
For $t\in\bT$ and $X\in\bar{L}^{0}$ define the (generalized) $\cF_{t}$-conditional expectation of $X$ by
$$
E[X|\cF_{t}]:=E[X^{+}|\cF_{t}]-E[X^{-}|\cF_{t}],
$$
where $X^{+}=(X\vee 0)$ and $X^{-}=(-X\vee 0)$. See Appendix~\ref{A:cond} for some relevant properties of the generalized expectation.

For $X\in \bar{L}^{0}$ and $t\in\bT$, we will denote by $\Essinf_{t}X$ the unique (up to a set of probability zero), $\cF_{t}$-measurable random variable, such that
\begin{equation}\label{eq:essinf.b}
\essinf_{\omega\in A}X=\essinf_{\omega\in A}(\Essinf_{t}X),
\end{equation}
 for any $A\in\cF_{t}$. We call this random variable the \textit{$\cF_{t}$-conditional essential infimum of $X$}. Similarly, we define $\Esssup_{t}(X):=-\Essinf_{t}(-X)$, and we call it the \textit{$\cF_{t}$-conditional essential supremum of $X$}. Again, see Appendix~\ref{A:cond} for more details and some elementary properties of conditional essential infimum and supremum.

The next definition introduces the main object of this work.

\begin{definition}\label{def:LM-measure}
A family $\varphi=\{\varphi_{t}\}_{t\in\mathbb{T}}$ of maps $\varphi_{t}:\cX\to\bar{L}^{0}_{t}$ is a {\it Dynamic LM-measure} if $\varphi$ satisfies
\begin{enumerate}[1)]
\item {(Locality)}  $\1_{A}\varphi_{t}(X)=\1_{A}\varphi_{t}(\1_{A}\cdot_{t} X)$;
\item {(Monotonicity)} $X\leq Y \Rightarrow \varphi_{t}(X)\leq \varphi_{t}(Y)$;
\end{enumerate}
for any $t\in\bT$, $X,Y\in\cX$ and $A\in\cF_{t}$.
\end{definition}

It is well recognized that locality and monotonicity are two properties that must be satisfied by any reasonable dynamic measure of performance and/or measure of risk, and in fact are shared by most, if not all, of such measures studied in the literature. The monotonicity property is natural for any numerical representation of an order between the elements of $\cX$.  The locality property (also referred to as regularity, or zero-one law, or relevance) essentially means that the values of the LM-measure restricted to a set $A\in\cF$ remain invariant with respect to the values of the arguments outside of the same set $A\in\cF$; in particular, the events that will not happen in the future do not affect the value of the measure today.

\begin{remark}
While in most of the literature the axiom of locality is not stated directly, it is very often implied by other assumptions.
For example, if $\cX=L^{\infty}$, then {\it monotonicity} and {\it cash-additivity}  imply locality (cf. \cite[Proposition 2.2.4]{Pitera2014PhD}). Similarly, any {\it convex} (or {\it concave}) map is also local (cf. \cite{DetlefsenScandolo2005}).
It is also worth mentioning that locality is  strongly related to time consistency. In fact, in some papers locality is considered as a part of the time consistency property discussed below (see e.g. \cite{KupperSchachermayer2009}).
\end{remark}

In this paper, we only consider dynamic LM-measures $\varphi$, such that
\begin{equation}\label{eq:normalization}
0\in \varphi_{t}[\cX],
\end{equation}
for any $t\in\bT$. We impose this (technical) assumption  to ensure that the maps $\varphi_t$ that we consider are not degenerate in the sense that they are not taking infinite values for all $X\in \cX$ on some set $A_t\in \cF_t$ of positive probability, for any $t\in\bT$; in the literature, sometimes such maps are referred to as {\it proper} \cite{Kaina2009}. If this is the case, then there  exists a family $\{Y_{t}\}_{t\in\bT}$, where $Y_{t}\in \cX$, such that $\varphi_{t}(Y_t)\in L^{0}_{t}$ for any $t\in\bT$, and so we can  consider  maps $\tilde{\varphi}$ given by $\tilde{\varphi}_t(\cdot):=\varphi_{t}(\cdot)-\varphi_{t}(Y_t)$, that satisfy assumption \eqref{eq:normalization} and preserve the same order as the maps $\varphi_{t}$ do. Typically, in the risk measure framework, one assumes that $\varphi_t(0)=0$, which implies \eqref{eq:normalization}. However, here we cannot assume that $\varphi_t(0)=0$, as we will also deal with dynamic performance measures for which $\varphi_t(0)=\infty$.

Finally, let us note that  in the literature, traditionally the dynamic risk measures are monotone decreasing. On the other hand, the measures of performance are monotone increasing. In view of condition 2) in Definition~\ref{def:LM-measure}, whenever our LM-measure corresponds to a dynamic risk measure, it needs to be understood as the negative of that risk measure. In such cases, in order to avoid confusion, we refer to the respective LM-measure as to  {\it dynamic (monetary) utility measure} rather than as {\it dynamic (monetary) risk measure}. See Appendix~\ref{S:families} for details.

\section{Approaches to time consistent assessment of preferences}\label{sec:TimeCons}

In this section, we present a brief survey of approaches to time consistent assessment of preferences, or to time consistency---for short, that were studied in the literature. As discussed in the Introduction, time consistency is studied via numerical representations of preferences. Various numerical representations will be surveyed below, and discussed in the context of dynamic LM-measures.

To streamline the presentation, we focus our attention on the case of random variables, that is  $\cX=L^{p}$, for $p\in\set{0,1,\infty}$.\footnote{{Most} of the concepts discussed in this Section can be modified to deal with the case of stochastic processes, as we will do in Section~\ref{s:tc.sp}.} Usually, the risk measures and the performance measures are studied on spaces smaller than $L^{0}$, such as $L^p,\, p\in[1,\infty].$ This is motivated by the aim to obtain so called robust representation of such measures {(see Appendix~\ref{S:families})}, since a certain topological structure is required for that (cf. Remark~\ref{rem:robustExt}). On the other hand, {\it time consistency} refers only to consistency of measurements in time, where no particular topological structure is needed, and thus most of the results obtained here hold true for $p=0$.

In Section \ref{s:GA}, we outline two generic approaches to time consistent assessment of preferences: an approach based on update rules and an approach based on benchmark families. These two approaches are generic in the sense that nearly all types of time consistency can be represented within these two approaches. On the contrary, the approaches outlined in Section \ref{s:SA} are specific. That is to say, those approaches are suited only for specific types of time consistency, specific classes of dynamic LM-measures, specific spaces, etc.

\subsection{Generic Approaches}\label{s:GA}

In this section, we outline two concepts that underlie the generic approaches to time consistent assessment of preferences: the update rules and the benchmark families. It will be seen that different types of time consistency can be characterized in terms of these concepts.

\subsubsection{Update rules}\label{sec:update.rules}

The approach to time consistency using update rules was developed in~\cite{BCP2014a}.
An update rule is a tool that is applied to preference levels, and used for relating  assessments of preferences done using a dynamic LM-measure at different times.

\begin{definition}\label{def:UMST}
A family $\mu=\{\mu_{t,s}:\, t,s\in\bT,\, t<s\}$ of maps $\mu_{t,s}:\bar{L}^{0}_{s}\to\bar{L}^{0}_{t}$ is called an {\it update rule} if $\mu$ satisfies the following conditions:
\begin{enumerate}[1)]
\item (Locality) $\1_{A}\mu_{t,s}(m)=\1_{A}\mu_{t,s}(\1_{A}m)$;
\item (Monotonicity) if $m\geq m'$, then  $\mu_{t,s}(m)\geq \mu_{t,s}(m')$;
\end{enumerate}
for any $s>t$, $A\in\cF_{t}$, and $m,m'\in\bar{L}^{0}_{s}$.
\end{definition}
Next, we give a definition of time consistency in terms of update rules.
\begin{definition}\label{def:tmbdotc}
Let $\mu$ be an update rule. We say that the dynamic LM-measure $\varphi$ is {\it $\mu$-acceptance (resp. $\mu$-rejection) time consistent} if
\begin{equation}\label{eq:atc}
\varphi_{s}(X)\geq m_{s} \quad(\textrm{resp. } \leq)\quad \Longrightarrow\quad \varphi_{t}(X)\geq \mu_{t,s}(m_{s})\quad(\textrm{resp. } \leq),
\end{equation}
for all $s>t$, $s,t\in\bT$, $X\in \cX$, and $m_{s}\in \bar{L}^{0}_{s}$. If property \eqref{eq:atc} is satisfied for $s=t+1$, $t\in\bT'$, then we say that $\varphi$ is {\it one-step $\mu$-acceptance (resp. one-step $\mu$-rejection) time consistent}.
\end{definition}

We see that $m_s$ and $\mu_{t,s}(m_s)$ serve as  benchmarks to which the measurements of $\varphi_{s}(X)$ and $\varphi_{t}(X)$ are compared, respectively. Thus, the interpretation of acceptance time consistency is straightforward: if $X\in\cX$ is accepted at some future time $s\in\bT$, at least at level $m_s$, then today, at time $t\in\bT$, it is accepted at least at level $\mu_{t,s}(m_s)$. Similar reasoning holds for the rejection time consistency. Essentially, the update rule $\mu$ converts the preference levels at time $s$ to the preference levels at time $t$.

We started our survey of time consistency with Definition \ref{def:tmbdotc} since, as we will demonstrate below,  this concept of time consistency covers various cases of time consistency for risk and performance measures that can be found in the existing literature. In particular, it allows  to establish important connections between different types of time consistency. The time consistency property of an LM-measure, in general, depends on the choice of the updated rule; we refer to Section~\ref{s:tc.rv} for an in-depth discussion.

It is useful to observe that the time consistency property given in terms of update rules can be equivalently  formulated as a version of the dynamic programming principle (see \cite[Proposition 3.6]{BCP2014a}):  $\varphi$ is $\mu$-acceptance (resp. $\mu$-rejection) time consistent if and only if
\begin{equation}\label{eq:accepTimeConsAlt}
\varphi_{t}(X)\geq \mu_{t,s}(\varphi_{s}(X))\quad (\textrm{resp. } \leq),
\end{equation}
for any $X\in\cX$ and $s,t\in\bT$, such that $s>t$. The interpretation of \eqref{eq:accepTimeConsAlt} is as follows: if the numerical assessment of preferences  about $X$ is given in terms of a dynamic LM-measure $\varphi$, then this measure is $\mu$-acceptance  time consistent if and only if the numerical assessment of preferences  about $X$ done at time $t$ is greater than the value of the measurement done at any future time $s>t$ and updated at time $t$ via $\mu_{t,s}$.
The analogous interpretation applies to the ejection time consistency.

Next, we define two interesting and important  classes of update rules.
\begin{definition}
Let $\mu$ be an update rule. We say that $\mu$ is
\begin{enumerate}[1)]
\item {\it $s$-invariant}, if there exists a family $\{\mu_{t}\}_{t\in\bT}$ of maps $\mu_{t}:\bar{L}^{0}\to\bar{L}^{0}_{t}$, such that $\mu_{t,s}(m_s)=\mu_{t}(m_s)$ for any $s,t\in\bT$, $s>t$, and  $m_s\in\bar{L}^{0}_{s}$;
\item {\it projective}, if it is $s$-invariant and $\mu_{t}(m_{t})=m_{t}$, for any $t\in\bT$, and $m_{t}\in\bar{L}^{0}_{t}$.
\end{enumerate}
\end{definition}

\begin{remark}\label{rem:UMnotation}
If an update rule $\mu$ is $s$-invariant, then  it is enough to consider only the corresponding family $\{\mu_{t}\}_{t\in\bT}$. Hence, with slight abuse of notation, we write $\mu=\{\mu_{t}\}_{t\in\bT}$ and call it an update rule as well.
\end{remark}

\begin{example}
The families  $\mu^1=\{\mu^1_t\}_{t\in\bT}$ and $\mu^2=\{\mu^2_t\}_{t\in\bT}$ given by
\[
\mu^1_t(m)=E[m|\cF_{t}],\quad \textrm{and}\quad \mu^2_t(m)=\Essinf_t m,\quad m\in \bar{L}^0,
\]
are projective update rules.
It will be shown in Example~\ref{ex:2} that there is  a dynamic LM-measure that is $\mu^2$--time consistent but not $\mu^1$--time consistent.
\end{example}

\subsubsection{Benchmark families}

The approach to time consistency based on families of benchmark sets was initiated by  \cite{Tutsch2008}, where the author applied this approach in the context of dynamic risk measures. Essentially, a benchmark family is a collection of subsets  of  $\cX$ that contain reference or test objects.
The idea of time consistency in this context,  is that the preferences about  objects of interest must compare in a consistent way to the preferences about the reference objects.

\begin{definition} \label{def:TC-benchSet} \mbox{}

(i) A family $\cY=\{\cY_{t}\}_{t\in\bT}$ of sets $\cY_{t}\subseteq \cX$ is a {\it benchmark family} if
\[
0\in\cY_{t}\quad \textrm{ and }\quad \cY_t+\bR=\cY_t,
\]
for any $t\in\bT$.

(ii) A dynamic LM-measure $\varphi$ is {\it acceptance (resp. rejection) time consistent with respect to the benchmark family $\cY$}, if
\begin{equation}\label{eq:benchmark2}
\varphi_{s}(X)\geq \varphi_{s}(Y)\quad (resp. \leq)\quad \Longrightarrow\quad \varphi_{t}(X)\geq \varphi_{t}(Y)\quad (resp. \leq),
\end{equation}
for all $s\geq t$, $X\in \cX$, and $Y\in\cY_{s}$.
\end{definition}
Informally, the ``degree'' of time consistency with respect to $\cY$ is measured by the size of $\cY$.
Thus, the larger the sets $\cY_{s}$ are, for each $s\in\bT$, the stronger the degree of time consistency of $\varphi$.

\begin{example}
The families of sets  $\cY^1=\{\cY^1_t\}_{t\in\bT}$ and $\cY^2=\{\cY^2_t\}_{t\in\bT}$ given by
\[
\cY^1_t=\bR\quad \textrm{and}\quad \cY^2_t=\cX,
\]
are benchmark families. They relate to weak and strong types of time consistency, as will be discussed later on.
\end{example}

For future reference, we recall from \cite[Proof of Proposition 3.9]{BCP2014a} that  $\varphi$ is acceptance (resp. rejection) time consistent with respect to $\cY$, if and only if $\varphi$ is acceptance (resp. rejection) time consistent with respect to the benchmark family $\widehat \cY$ given by
\begin{equation}\label{eq:bench.local}
\widehat{\cY_{t}}:=\{Y\in \cX: Y=\1_A Y_1 +\1_{A^c} Y_2, \textrm{ for some } Y_1,Y_2\in\cY_{t} \textrm{ and } A\in\cF_t\}.
\end{equation}

\subsubsection{Relation between update rule approach and the benchmark approach}

The difference between the update rule approach and the benchmark family approach is  that the preference levels are chosen differently. Specifically, in the former approach, the preference level at time $s$ is chosen as any $m_s\in \bar{L}^{0}_{s}$, and then updated to the preference level at time $t$, using an update rule. In the latter approach, the preference levels at both times $s$ and $t$ are taken as  $\varphi_s(Y)$ and $\varphi_t(Y)$, respectively,  for any reference object $Y\in \cY_s$, where $\cY_s$ is an element of the benchmark family $\cY$.

These two approaches are strongly related to each other. Indeed, for any LM-measure $\varphi$ and for any benchmark family $\cY$, one can construct an update rule $\mu$ such that $\varphi$  is time consistent with respect to $\cY$ if and only if it is $\mu$-time consistent.

For example, in case of acceptance time consistency of $\varphi$ with respect to $\cY$, using the locality of $\varphi$, it is easy to note that \eqref{eq:benchmark2} is equivalent to
\[
\varphi_{t}(X)\geq \esssup_{A\in\cF_{t}}\Big[\1_{A}\esssup_{Y\in \cY^{-}_{A,s}(\varphi_s(X))}\varphi_{t}(Y)+\1_{A^{c}}(-\infty)\Big],
\]
where $\cY^{-}_{A,s}(m_s):=\{Y\in\widehat{\cY}_s: \1_A m_s\geq \1_A\varphi_s(Y)\}$ and $\widehat{\cY}=\{\widehat{\cY}_s \}_{s\in\bT}$ is defined in \eqref{eq:bench.local}. Consequently, setting
\[
\widetilde{\mu}_{t,s}(m_s):=\esssup_{A\in\cF_{t}}\Big[\1_{A}\esssup_{Y\in \cY^{-}_{A,s}(m_s)}\varphi_{t}(Y)+\1_{A^{c}}(-\infty)\Big],
\]
and using \eqref{eq:accepTimeConsAlt}, we deduce that $\varphi$ satisfies \eqref{eq:benchmark2} if and only if $\varphi$ is time consistent with respect to the update rule $\widetilde \mu_{t,s}$ (see \cite[Proposition 3.9]{BCP2014a} for details). The analogous argument works for rejection time consistency.

Generally speaking, the converse implication does not hold true; the notion of time consistency given in terms of update rules is more general. For example, time consistency of a dynamic coherent acceptability index cannot be expressed  in terms of a single benchmark family.

\subsection{Idiosyncratic Approaches}\label{s:SA}
Each such approach to time consistency of a given LM-measure exploits the idiosyncratic properties of this LM-measure, which are not necessarily shared by other LM measures, and typically is suited only for a specific subclass of dynamic LM-measures.
For example, in case of dynamic convex or monetary risk measures the time consistency can be characterized in terms of the relevant properties of associated acceptance sets and/or the dynamics of the penalty functions and/or the rectangular property of the families of probability measures. These idiosyncratic approaches, and the relevant references, were mentioned and briefly discussed in Section~\ref{s:litRev}. Detailed analysis of each of these approaches is beyond the scope of this survey.

\section{Time consistency for random variables}\label{s:tc.rv}

In this Section, we survey the time consistency of LM-measures applied to random variables. Accordingly, we assume that $\cX=L^{p}$,  for a fixed $p\in\{0,1,\infty\}$. We proceed with the discussion of various related types of time consistency, without much reference to the existing literature. Such references are provided in Section~\ref{s:litRev}.

\subsection{Weak time consistency}
The main idea behind this type of time consistency is that if ``tomorrow'', say at time $s$, we accept $X\in L^p$ at level $\varphi_{s}(X)$, then ``today'', say at time  $t$, we would accept  $X$ at any level less than or equal to $\varphi_{s}(X)$, adjusted by the information $\cF_t$ available at time $t$. Similarly, if tomorrow we reject $X$ at level $\varphi_{s}(X)$, then today, we should also reject $X$ at any level greater than or equal to $\varphi_{s}(X)$, adapted to the information $\cF_t$.

\begin{definition}\label{type.of.cons.weak}
A dynamic LM-measure $\varphi$ is {\it weakly acceptance (resp. weakly rejection) time consistent} if
\[
\varphi_{t}(X)\geq \Essinf_{t}\varphi_{s}(X),\quad (resp.\quad \varphi_{t}(X)\leq \Esssup_{t}\varphi_{s}(X)\,)
\]
for any $X\in L^p$ and $s,t\in\bT$, such that $s>t$.
\end{definition}

Propositions~\ref{pr:weak} and~\ref{pr:weak2} provide some characterizations of weak acceptance time consistency.
\begin{proposition}\label{pr:weak}
Let $\varphi$ be a dynamic LM-measure on $L^{p}$. The following properties are equivalent:
\begin{enumerate}[1)]
\item $\varphi$ is weakly acceptance time consistent.
\item $\varphi$ is $\mu$-acceptance time consistent, where $\mu$ is a projective update rule, given by
\[
\mu_{t}(m)=\Essinf_{t}m.
\]
\item The following inequality is satisfied
\begin{equation}\label{eq:weak.radon}
\varphi_{t}(X)\geq \essinf_{Q\in\cM_t(P)}E_{Q}[\varphi_{s}(X)|\mathcal{F}_{t}],
\end{equation}
for any $X\in L^p$,  $s,t\in\bT$, $s>t$.
\item For any $X\in L^p$, $s,t\in\bT$, $s>t$, and $m_{t}\in \bar{L}^{0}_{t}$, it holds that
\[
\varphi_{s}(X)\geq m_{t} \Rightarrow \varphi_{t}(X)\geq m_{t}.
\]
\end{enumerate}
Similar results hold true for weak rejection time consistency.
\end{proposition}
For the proof of the equivalence between 1), 2), and 4), see \cite[Proposition 4.3]{BCP2014a}. Regarding 3), note that any measure $Q\in\cM_t(P)$ may be expressed in terms of a Radon-Nikodym derivative with respect to measure $P.$ In other words, instead of \eqref{eq:weak.radon}, we may write
\[
\varphi_{t}(X) \geq \essinf_{Z\in P_{t}}E[Z\varphi_{s}(X)|\mathcal{F}_{t}],
\]
where $P_{t}:=\{Z\in L^{1} \mid Z\geq 0,\ E[Z|\mathcal{F}_{t}]=1\}$. Thus, one can show equivalence between 1) and 3) noting that for any $m\in \bar{L}^{0}$ we get $\Essinf_{t}m= \essinf_{Z\in P_{t}}E[Zm|\mathcal{F}_{t}]$. See \cite[Proposition 4.4]{BCP2014a} for the proof.

It is worth mentioning that Property 4) in Proposition~\ref{pr:weak} was suggested as the notion of (weak) acceptance and (weak) rejection time consistency in the context of scale invariant measures, called acceptability indices (cf.~\cite{BiaginiBion-Nadal2012,BCZ2010}).

Usually, the weak time consistency is considered for dynamic monetary risk measures on $L^{\infty}$ (cf.~\cite{AcciaioPenner2010} and references therein). This case lends itself to even more characterizations of this property.

\begin{proposition}\label{pr:weak2}
Let $\varphi$ be a {representable dynamic monetary utility measure}\,\footnote{See section \ref{S:families} for details.} on $L^{\infty}$. The following properties are equivalent:
\begin{enumerate}[1)]
\item $\varphi$ is weakly acceptance time consistent.
\item $\varphi$ is acceptance time consistent with respect to $\{\cY_t\}_{t\in\bT}$, where $\cY_t=\bR$.
\item For any $X\in L^p$ and $s,t\in\bT$, $s>t$,
\begin{equation}\label{eq:classicWeak}
\varphi_{s}(X)\geq 0 \Rightarrow \varphi_{t}(X)\geq 0.
\end{equation}
\item $\cA_{t+1}\subseteq \cA_{t}$, for any $t\in\bT$, such that $t<T$.
\item For any $Q\in \cM(P)$ and $t\in\bT$, such that $t<T$,
\[
\alpha^{\min}_{t}(Q)\geq E_{Q}[\alpha^{\min}_{t+1}(Q)\,|\, \cF_{t}],
\]
where $\alpha^{\min}$ is the  minimal penalty function in the robust representation of $\varphi$.
\end{enumerate}
Analogous  results are obtained for weak rejection time consistency.
 \end{proposition}

We note that equivalence of properties 1), 2), and 3) also holds true in the case of $\cX=L^{0}$, and not only for representable, but for any dynamic monetary utility measure;  for the proof, see \cite[Proposition 4.3]{BCP2014a}.
Property 4) is a characterisation of weak time consistency in terms of acceptance sets, and property~5) gives a characterisation in terms of the supermartingale property of the penalty function.
For the proof of the equivalence of 3), 4), and 5), see~\cite[Proposition 33]{AcciaioPenner2010}.

The next result shows that weak time consistency is indeed one of the weakest forms of time consistency, in the sense that the weak time consistency is implied by any time consistency generated by a projective update rule;
 we refer to \cite[Proposition 4.5]{BCP2014a} for the  proof.

\begin{proposition}\label{pr:UDMprop}
Let $\varphi$  be a dynamic LM-measure on $L^p$, and let $\mu$ be a projective update rule. If $\varphi$ is $\mu$-acceptance (resp. $\mu$-rejection) time consistent, then $\varphi$ is weakly acceptance (resp. weakly rejection) time consistent.
\end{proposition}

\begin{remark}\label{cor:weak.pre}
An important feature of the weak  time consistency   is its invariance with respect to monotone transformations.
Specifically, let $g:\bar{\bR}\to\bar{\bR}$ be a strictly increasing function and let $\varphi$ be a weakly acceptance/rejection  time consistent dynamic LM-measure.
Then, $\{g\circ\varphi_{t}\}_{t\in\bT}$ is also a weakly acceptance/rejection time consistent dynamic LM-measure.
\end{remark}

\begin{remark} In the case of general LM-measures, the weak time consistency may not be characterized as in 2) of Proposition~\ref{pr:weak2}.
For example, if $\varphi$  is a (normalized) acceptability index, then  $\varphi_t(\bR)=\{0,\infty\}$, for $t\in\bT$, which does not agree with 4) in Proposition~\ref{pr:weak}.
\end{remark}

\subsection{Strong time consistency}\label{S:strong.rv}

As already stated in the Introduction, the origins of the strong form of time consistency can be traced to \cite{Koopmans1960}. Historically, this is the first and the most extensively studied form of time consistency in dynamic risk measures literature. It is fair to mention, that this form of time consistency also appears in the insurance literature, as the iterative property, and it is related to the mean value principle  \cite{Gerber1974,GoovaertsDeVylder1979}.

We start with the definition of strong time consistency.

\begin{definition}\label{type.of.cons.strong}
Let $\varphi$ be a dynamic LM-measure on $L^p$.
Then, $\varphi$ is said to be
{\it strongly time consistent} if
\begin{equation}\label{eq:strongT}
\varphi_{s}(X)= \varphi_{s}(Y) \quad\Longrightarrow \quad\varphi_{t}(X)= \varphi_{t}(Y),
\end{equation}
for any $X,Y\in L^p$ and $s,t\in\bT$, such that $s>t$.
\end{definition}

Strong time consistency gains its popularity and importance due to its equivalence  to the dynamic programming principle. This equivalence, as well as other  characterisations of strong time consistency, are the subject of the following two propositions.

\begin{proposition}\label{pr:strong}
Let $\varphi$ be a dynamic LM-measure on $L^p$. The following properties are equivalent:
\begin{enumerate}[1)]
\item $\varphi$ is strongly time consistent.
\item There exists an update rule $\mu$ such that $\varphi$ is both $\mu$-acceptance and $\mu$-rejection time consistent.
\item $\varphi$ is acceptance time consistent with respect to $\{\cY_t\}_{t\in\bT}$, where $\cY_t=L^p$.
\item There exists an update rule $\mu$ such that for any $X\in L^p$, $s,t\in\bT$, $s>t$,
\begin{equation}\label{eq:DP.strong}
\mu_{t,s}(\varphi_{s}(X))=\varphi_{t}(X).
\end{equation}
\item There exists a one-step update rule $\mu$ such that for any $X\in L^p$, $t\in\bT$, $t<T$,
\[
\mu_{t,t+1}(\varphi_{t+1}(X))=\varphi_{t}(X).
\]
\end{enumerate}
\end{proposition}

See Appendix~\ref{A:proofs} for the proof of Proposition~\ref{pr:strong}. Property 4) in this proposition is referred to as Bellman's principle or the dynamic programming principle. Also, note that 5) implies that any strongly time consistent dynamic LM-measure can be constructed using a backward recursion starting from $\varphi_T:=\varrho$, where $\varrho$ is an LM-measure.   {See  \cite{CheriditoKupper2006} where the recursive construction for dynamic risk measures is discussed in details.}

An important, and frequently studied, type of strong time consistency is the strong time consistency for dynamic monetary risk measures on $L^{\infty}$ (cf.~\cite{AcciaioPenner2010} and references therein). As the next result shows, there are more equivalences that are valid in this case.

\begin{proposition}\label{pr:strong2}
Let $\varphi$ be a {representable dynamic monetary utility measure} on $L^{\infty}$. The following properties are equivalent:
\begin{enumerate}[1)]
\item $\varphi$ is strongly time consistent.
\item  $\varphi$ is recursive, i.e., for any $X\in L^p$, $s,t\in\bT$,  $s>t$,
\[
\varphi_{t}(X)= \varphi_{t}(\varphi_{s}(X)).
\]
\item $\cA_t=\cA_{t,s}+\cA_{s}$, for all $t,s\in\bT$, $s>t$.
\item For any $Q\in \cM(P)$, $t,s\in\bT$, $s>t$,
\[
\alpha_t^{\min}(Q)=\alpha_{t,s}^{\min}(Q)+E_{Q}[\alpha^{\min}_{s}(Q)\,|\, \cF_{t}].
\]
\item For any $X\in L^p$, $Q\in \cM(P)$, $s,t\in\bT$, $s>t$,
\[
\varphi_{t}(X)-\alpha^{\min}_{t}(Q)\leq E_{Q}[\varphi_{s}(X)-\alpha^{\min}_{s}(Q)\,|\, \cF_{t}].
\]

\end{enumerate}
\end{proposition}

\noindent
For the proof see, for instance, \cite[Proposition 14]{AcciaioPenner2010}.

\begin{remark}\label{rem:MidNotWeak} (i) In general, for dynamic LM-measures, the strong time consistency does not imply either the weak acceptance or  weak rejection time consistency.
Indeed, let us consider $\varphi=\{\varphi_{t}\}_{t\in\bT}$, such that $\varphi_{t}(X)=t$ (resp. $\varphi_{t}(X)=-t$) for all $X\in L^{0}$.
Since $\varphi_{t}(0)=t\not\geq \essinf_{t}\varphi_{s}(0)=s$ (resp. $-t\not\leq -s$), for $s>t$, we conclude that $\varphi$ is not weakly acceptance (resp. weakly rejection) time consistent. However, since $\varphi_{t}(X)=\varphi_{t}(\varphi_{s}(X))$ for any $X\in L^{0}$, then $\varphi$ is strongly time consistent. We note, that if the update rule in Definition~\ref{type.of.cons.strong} is projective, as it is usually the case for dynamic monetary risk measures, then, due to Proposition~\ref{pr:UDMprop}, the strong time consistency implies the weak time consistency.\\
(ii) It is worth mentioning that, in principle,  strong time consistency is not suited for acceptability indices \cite{BCZ2010,BCC2014,ChernyMadan2009}.
 Let $\varphi$  be a scale invariant dynamic LM-measure, and let $A\in\cF_{s}$ be such that $P[A]=1/2$, for some $s>0$, $s\in\bT$. Additionally, assume that $\cF_0$ is trivial.
 We consider the sequence of random variables  $X_n=n\1_{A}-\1_{A^{c}}$, $n\in\bN$.
 By locality and scale invariance of $\varphi$, we have that $\varphi_s(X_n)=\varphi_s(X_1)$, for $n\in\bN$.
 If $\varphi$ is strongly time consistent, then we also have that $\varphi_0(X_n)=\varphi_0(X_1)$, $n\in\bN$. On the other hand, any reasonable measure of performance should assess $X_n$ at the higher level as $n$ increases, which contradicts the fact that $\varphi_0(X_n)$ is a constant sequence.
 \end{remark}

\subsection{Robust expectations, submartingales, and supermartingales}
The concept of a projective update rule is  connected with the concept of the (conditional) nonlinear expectation (see, for instance, \cite{Peng1997} for the definition and properties of nonlinear expectation).
In \cite{RosazzaGianin2006,Peng2004}, the authors established a link between nonlinear expectations and dynamic risk measures.
 One particularly important example of an projective update rule is the standard conditional expectation operator.
 Time consistency in $L^{\infty}$ framework, defined in terms of conditional expectation,  was studied in  \cite[Section~5]{DetlefsenScandolo2005} and associated with the super(sub)martingale property.

The next result introduces a general class of updates rules that are generated by conditional expectations and determining families of sets. First, we recall the concept of the determining family of sets (see, for instance, \cite{ChernyWVAR2006} for more details).

For each $t\in\bT$ define
 \[
 \cP_{t}:=\{Z\in L^{1} \mid Z\geq 0,\ E[Z|\mathcal{F}_{t}]=1\}.
 \]
A family of sets $\cD=\{\cD_{t}\}_{t\in\bT}$ is a {\it determining family} if for any $t\in\bT$, the  set $\cD_{t}$ satisfies the following properties: $\cD_{t}\neq \emptyset$,  $\cD_{t}\subseteq \cP_{t}$, it is $L^{1}$-closed, $\cF_{t}$-convex\footnote{By $\cF_{t}$-convex we mean that for any $Z_{1},Z_{2}\in \cD_{t}$, and $\lambda\in L^{0}_{t}$ such that $0\leq\lambda\leq 1$ we get $\lambda Z_{1}+(1-\lambda)Z_{2}\in \cD_{t}$.}, and uniformly integrable.

\begin{proposition}\label{pr:coh.upd}
Let $\cD$ be a determining family of sets, and let $\varphi$ be a dynamic LM-measure.
Consider the family of maps $\phi=\{\phi_{t}\}_{t\in\bT}$, $\phi_{t}:\bar{L}^{0}\to\bar{L}^{0}_{t}$, given by the following robust expectations\footnote{The term robust is inspired by robust representations of risk measures.}
\begin{equation}\label{eq:robust}
\phi_{t}(m)=\essinf_{Z\in \cD_{t}}E[Zm|\cF_{t}].
\end{equation}
Then,
\begin{enumerate}[1)]
\item the family $\phi$ is a projective update rule;
\item if $\varphi$ is $\phi$-acceptance time consistent, then  $\{g\circ\varphi_{t}\}_{t\in \bT}$ is also $\phi$-acceptance time consistent, for any  increasing and concave function $g:\bar{\bR}\to\bR$.
\end{enumerate}
\end{proposition}

\begin{remark}
Classical (static) coherent risk measures defined on $L^\infty$ admit robust representation of the form \eqref{eq:crmStatic}
for some set of probability measures $\cQ$. It is known that the set $\cQ$ might not be unique.
Consequently, there may exist multiple extensions of $\rho$ to a map defined on $\bar L^0$ (see Appendix~\ref{S:extensions} for the concept of the extension).  Nevertheless, as in~\cite{ChernyWVAR2006}, one can consider the maximal set $\cD$ called \textit{determining set} of a risk measure, which guarantees the uniqueness of such extension.
The family of maps defined in \eqref{eq:robust} is an example of a family of such extensions.
Consequently, we see that the coherent risk measures constitute a good starting point for generation of update rules.
\end{remark}

For the proof of Proposition~\ref{pr:coh.upd}, see Appendix~\ref{A:proofs}. The counterpart of Proposition~\ref{pr:coh.upd} for rejection time consistency is obtained by taking $\esssup$ instead of $\essinf$ in \eqref{eq:robust}, and assuming that $g$ is convex.

In the particular case of determining family with $\cD_t=\set{1}$, for any $t\in\bT$,  the projective update rule takes the form  $\mu_{t}(m)=E[m|\cF_{t}]$,  $m\in \bar{L}^{0}$. This is an important case, as it produces the concept of supermartingale and submartingale time consistency.

\begin{definition}\label{type.of.cons.expectation}
Let $\varphi$ be a dynamic LM-measure on $L^p$. We say that $\varphi$ is \textit{supermartingale (resp. submartingale) time consistent} if
\[
\varphi_t(X)\geq E[\varphi_s(X) | \cF_t],\quad\quad (\textrm{resp. } \leq)
\]
for any $X\in L^p$ and $t,s\in\bT$, $s>t$.
\end{definition}

\begin{remark}
  \label{rem:subThenWeak} (i)  Note that any dynamic LM-measure that is $\phi$-acceptance time consistent, where $\phi$ is given in \eqref{eq:robust}, is also weakly acceptance time consistent, as $\phi$ is projective. In particular, any supermartingale time consistent LM-measure is also weakly acceptance time consistent. A similar statement holds true for rejection time consistency.\\
  (ii) As mentioned in~\cite{BCP2014a}, the idea of update rules might be used to weight the preferences. {Intuitively speaking, the risk of loss in the far future might be more preferred than the imminent risk of loss.} This idea was used in \cite{Cherny2010}. For example, the update rule $\mu$ of the form
\begin{equation}\label{eq:ur:exp3}
\mu_{t,s}(m,X)=\left\{
\begin{array}{ll}
\alpha^{s-t} E[m|\cF_{t}] & \textrm{on } \{E[m|\cF_{t}] \geq 0\},\\
\alpha^{t-s} E[m|\cF_{t}] &  \textrm{on } \{E[m|\cF_{t}] < 0\}.
\end{array}\right.
\end{equation}
for a fixed $\alpha\in (0,1)$ would achieve this goal.
\end{remark}

\subsection{Other types of time consistency}\label{S:other.forms}
The weak, strong, and super/sub-martingale forms of time consistency have attracted the most attention in the existing literature. In this section, we present other forms of time consistency that have been studied.

\subsubsection{Middle time consistency}
The notion of middle time consistency was originally formulated for dynamic monetary risk measures on $L^{\infty}$ (cf.~\cite{AcciaioPenner2010}). The main idea is to replace the equality in \eqref{eq:strongT} by an inequality. The term {\it middle acceptance} or {\it middle rejection} is used depending on the direction of the inequality.

\begin{definition}\label{type.of.cons.middle}
A dynamic LM-measure $\varphi$ on $L^p$ is  {\it middle acceptance (resp. middle rejection) time consistent} if
\[
\varphi_{s}(X)\geq \varphi_{s}(Y)\quad (\textrm{resp. } \leq)\quad \Longrightarrow\quad \varphi_{t}(X)\geq \varphi_{t}(Y)\quad (\textrm{resp. } \leq) ,
\]
for any $X\in L^p$, $s,t\in\bT$, $s>t$, and $Y\in L^p\cap {L}^{0}_{s}$.
\end{definition}

The middle acceptance (resp. middle rejection) time consistency is equivalent to the acceptance (resp. rejection) time consistency with respect to the benchmark family $\cY=\{\cY_{t}\}_{t\in\bT}$, given by $\cY_{t}=L^p\cap {L}^{0}_{t}$.
In the case of dynamic convex risk measures, other characterizations of middle acceptance time consistency are available, as the following proposition shows.

\begin{proposition}\label{pr:middle}
Let $\varphi$ be a representable dynamic monetary utility measure on $L^{\infty}$, which is continuous from above. The following properties are equivalent:
\begin{enumerate}[1)]
\item $\varphi$ is middle acceptance time consistent.
\item $\varphi$ is $\varphi^{-}$-acceptance time consistent.\footnote{See Appendix~\ref{S:extensions} for the definition of $\varphi^{-}$}
\item For any $X\in L^p$, $s,t\in\bT$,  $s>t$,
\begin{equation}\label{eq:DP.middle}
\varphi_{t}(X)\geq \varphi_{t}(\varphi_{s}(X)).
\end{equation}

\item For any $X\in L^p$ and $t\in\bT$, such that $t<T$,
\[
\varphi_{t+1}(X)-\varphi_{t}(X)\in\cR_{t,t+1}.
\]
\item For any $X\in\cR_t$ and $t\in\bT$, such that $t<T$,
\[
\varphi_{t+1}(X)\in\cR_{t}.
\]
\item
For any $t\in\bT$, such that $t<T$,
$ \cA_{t}\supseteq \cA_{t,t+1}+\cA_{t+1}$.

\item For any $Q\in\cM(P)$ and $t\in\bT$, such that $t<T$,
\[
\alpha_{t}^{\min}(Q) \geq \alpha_{t,t+1}^{\min}(Q)+E_{Q}[\alpha_{t+1}^{\min}(Q) | \cF_{t}].
\]
\item  For any $Q\in\cM(P)$ and $t\in\bT$, such that $t<T$,
\[
\varphi_{t}(X)\geq E_{Q}[\varphi_{t+1}(X)\mid\cF_{t}]+\alpha_{t,t+1}^{\min}(Q).
\]

\end{enumerate}
\end{proposition}
Since $\varphi^{-}$ is an LM-extenstion of $\varphi$, and $\varphi_{s}(Y)=Y$, for any $Y\in L^p \cap \bar{L}^{0}_{s}$, the equivalence between 1) and 2) is immediate.
For all other equivalences  see \cite[Section 4.2]{AcciaioPenner2010} and references therein.
Property 1) in Proposition~\ref{pr:middle} is sometimes called {\it prudence} (see \cite{Penner04}).


\subsubsection{Time consistency induced by LM-measure}\label{S:TCinduced}
It turns out that any dynamic LM-measure generates an update rule. Indeed, as the next result shows, any LM-extension of an LM-measure (see appendix \ref{S:extensions} for the definition of LM-extension) is an s-invariant update rule.

\begin{proposition}\label{pr:ext}
Any LM-extension $\widehat{\varphi}$ of a dynamic LM--measure $\varphi$ is an s-invariant update rule.
Moreover, $\widehat{\varphi}$ is projective if and only if $\varphi_{t}(X)=X$, for $t\in\bT$ and $X\in L^p\cap\bar{L}^{0}_{t}$.
\end{proposition}

\noindent The proof is deferred to Appendix~\ref{A:proofs}.

\smallskip
LM-extensions may be used to give stronger forms of strong and middle time consistency, that are especially well suited in the case of dynamic monetary risk measures.

Recall that for dynamic monetary risk measure $\varphi$ on $L^{\infty}$, strong time consistency is equivalent to the property that
\[
\varphi_{t}(X)= \varphi_{t}(\varphi_{s}(X)),
\]
for any $X\in \cX$, $s,t\in\bT$,  $s>t$.

However, if $\cX$ is larger than $L^{\infty}$, then this characterisation is problematic, as we might get $\varphi_{s}(X)\not\in \cX$. In this case, LM-extensions come in handy and one defines strong time consistency via the following equality,
\begin{equation}\label{eq:strong.strong}
\varphi_{t}(X)= \hat{\varphi}_{t}(\varphi_{s}(X)),\quad X\in\cX,\ s,t\in\bT,\  s>t,
\end{equation}
where  $\hat{\varphi}$ is an extension of $\varphi$ from $\cX$ to $\bar{L}^{0}$. Accordingly, we say that $\varphi$ is \textit{strongly$^{*}$ time consistent}, if there exists an LM-extension $\hat{\varphi}$, of $\varphi$, such that $\varphi$ is both $\hat{\varphi}$-acceptance and $\hat{\varphi}$-rejection time consistent.
\smallskip

\noindent Note that since $\hat{\varphi}$ is an update rule, the  strong$^{*}$ time consistency implies strong time consistency in the sense of Definition~\ref{type.of.cons.strong}. In general, the converse implication is not true; to see this, it is enough to consider strong time consistency for an update rule that is not s-invariant.

In the same fashion, we say that $\varphi$ is \textit{middle$^{*}$ acceptance time consistent}, if there exists an LM-extension of $\varphi$, say $\hat{\varphi}$, such that $\varphi$ is $\hat{\varphi}$-acceptance time consistent. In view of Proposition \ref{pr:upper.ext}, this is equivalent to saying that $\varphi$ is middle$^{*}$ acceptance time consistent if it is $\varphi^{-}$-acceptance time consistent. Likewise, to define \textit{middle$^*$ rejection time consistency} we use the mapping $\varphi^{+}$.

\subsection{Taxonomy of results}

For the convenience of the reader, in Flowchart~\ref{chart:rv} below, we summarize the results surveyed in Section \ref{s:tc.rv}. For transparency, we label (by circled numbers) each arrow (implication or equivalence) in the flowchart,  and we relate the labels to the relevant results, also providing comments on converse implications whenever appropriate.

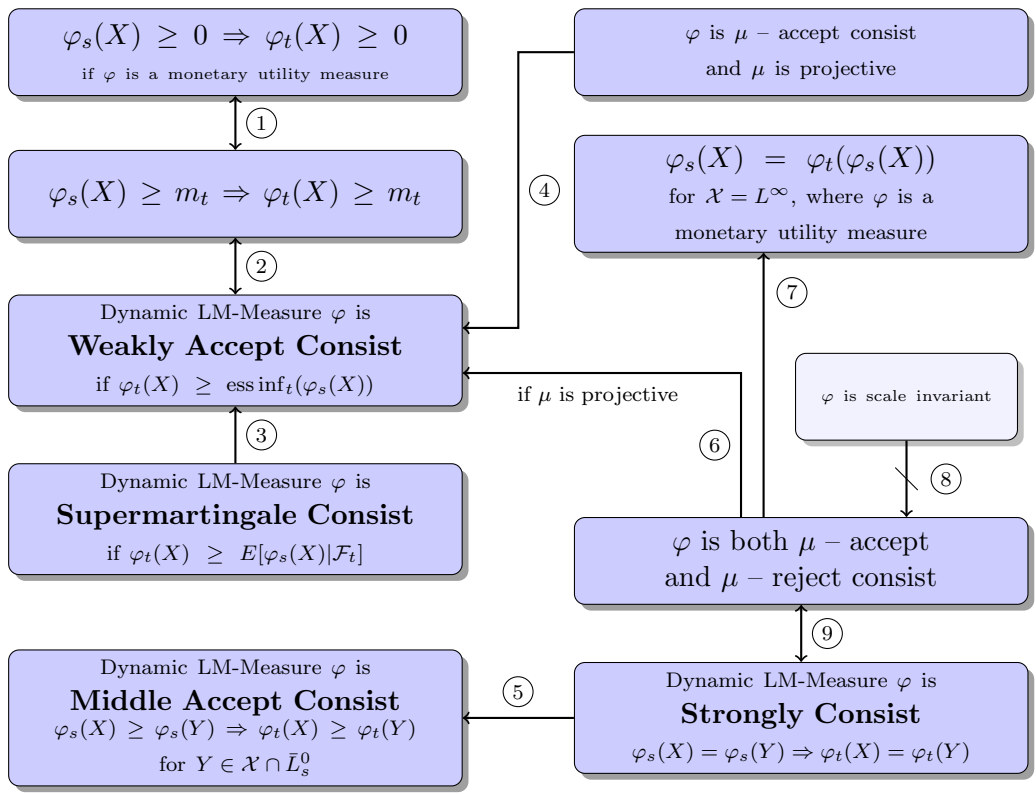
\begin{figure}
\caption{Summary of results for acceptance time consistency for random variables}
\begin{tikzpicture}[scale=1.0,transform shape]\label{chart:rv}
\path node (p0) [practica]
	{$\varphi_s(X)\geq 0 \Rightarrow \varphi_t(X)\geq 0$ \\
	\tiny{if $\varphi$ is  a monetary utility measure}};
\path  (p0.south) +(0.0,-1.3) node (p1) [practica]
	{$\varphi_s(X)\geq m_t \Rightarrow \varphi_t(X)\geq m_t$};
\path (p1.south)+(0.0,-1.5) node (p2) [practica]
	{ {\scriptsize Dynamic LM-Measure $\varphi$ is }  \\
	\textbf{Weakly Accept Consist }\\
	{\scriptsize if $\varphi_t(X)\geq \Essinf_t(\varphi_s(X))$} };
\path (p2.south)+(0.0,-1.5) node (p3) [practica]
	{ {\scriptsize Dynamic LM-Measure $\varphi$ is }  \\
	\textbf{Supermartingale Consist }\\
	{\scriptsize if $\varphi_t(X)\geq E[\varphi_s(X)|\mathcal{F}_t]$} };
\path (p3.south)+(0.0, -1.9) node (pp2) [practica]
	{
	 {\scriptsize Dynamic LM-Measure $\varphi$ is }  \\
	\textbf{Middle Accept Consist }\\
	\scriptsize{$\varphi_s(X)\geq \varphi_s(Y)\Rightarrow \varphi_t(X)\geq \varphi_t(Y)$}\\
	\mbox{{\scriptsize for $Y\in\cX\cap \bar{L}^{0}_{s}$}}
	};;
\path (p0.east)+(4.50, 0.0) node (p8) [practica]
	{{\scriptsize $\varphi$ is $\mu$ -- accept consist \\
	and $\mu$ is projective} };
\path (p1.east)+(4.50, 0.0) node (p18) [practica]
	{$\varphi_s(X)= \varphi_t(\varphi_s(X))$\\
	\mbox{{\scriptsize for $\cX=L^{\infty}$, where}}
	{\scriptsize $\varphi$ is a monetary utility measure}
	};
\path (p2.east)+(4.5, -2.8) node (p4) [practica]
	{ $\varphi$ is both $\mu$ -- accept and $\mu$ -- reject consist};

\path (pp2.east)+(4.5, 0.0) node (p5) [practica]
	{{\scriptsize Dynamic LM-Measure $\varphi$ is }  \\
	\textbf{Strongly Consist }\\
	{\scriptsize \mbox{$\varphi_s(X)= \varphi_s(Y)\Rightarrow \varphi_t(X) = \varphi_t(Y)$}}};
\path (p4.north)+(1.4, 1.6) node (p6) [practica0]
	{ {\tiny $\varphi$ is scale invariant}};
\path (p2.east)+(1.8,-0.6) node (p9) { {\scriptsize if $\mu$ is projective}};
\draw[<->,thick] (p0.south) --  node[anchor=west] {\circled{\scriptsize{1}}} (p1.north) ;
\draw[<->,thick] (p1.south) -- node[anchor=west] {\circled{\scriptsize{2}}} (p2.north);
\draw[<-,thick] (p2.south) -- node[anchor=west] {\circled{\scriptsize{3}}} (p3.north);
\draw[<-,thick] ($(p18.south)+(-0.5,0.0)$) -- node[pos=0.15, anchor=west] {\hspace{-0.1cm} \circled{ \scriptsize{7}}  } ($(p4.north)+(-0.5,0.0)$);
\draw[<->,thick] (p5.north) -- node[anchor=west] {\circled{\scriptsize{9}}} (p4.south);
\draw[<-,thick]  (pp2.east) -- node[anchor=south] {\circled{\scriptsize{5}}} (p5.west); 
\draw [->, strike-arrow,thick] (p6.south)  -- node[anchor=west] {\hspace{.05cm} \circled{\scriptsize{8}}} ($(p4.north)+(1.4, 0)$);
\path[->,draw,thick] (p8.west) -- ($(p8.west)!.5!(p0.east)$) -- node[anchor=west] {\circled{\scriptsize{4}}} ($($(p2.west)!($(p8.west)!.5!(p0.east)$)!(p2.east)$)+(0,0.3)$) --($(p2.east)+(0,0.3)$);
\path[->,draw,thick] ($(p4.north)+(-0.8,0)$) -- node[anchor=east] {\circled{\scriptsize{6}}} ($($(p2.west)!($(p4.north)+(-0.8,0)$)!(p2.east)$)+(0,-0.3)$) -- ($(p2.east)+(0,-0.3)$);
\end{tikzpicture}
\label{diagram.rv}
\end{figure}

\begin{enumerate}
  \item[\circled{\scriptsize{1}}] Proposition~\ref{pr:weak2}, 2)
  \item[\circled{\scriptsize{2}}] Proposition~\ref{pr:weak}, 4)
  \item[\circled{\scriptsize{3}}] Remark~\ref{rem:subThenWeak} and Proposition~\ref{pr:UDMprop}. The converse implication is not true in general, see Example~\ref{ex:5}.
  \item[\circled{\scriptsize{4}}] Proposition~\ref{pr:UDMprop}. Generally speaking, the converse implication is not true. See Example~\ref{ex:5}: the negative of Dynamic Entropic Risk Measure with $\gamma<0$ is weakly acceptance time consistent, but it is not supermaringale time consistent, i.e., it is not acceptance time consistent with respect to the projective update rule $\mu_t=E_t[m|\cF_t]$.

  \item[\circled{\scriptsize{5}}] Proposition~\ref{pr:strong}, 4). The converse implication is not true in general. For the counterexample,
  see \cite[Proposition 37]{AcciaioPenner2010}.

   \item[\circled{\scriptsize{6}}] Proposition~\ref{pr:UDMprop}, and see also \circled{\scriptsize{4}}. In general, strong time consistency does not imply weak acceptance time consistency, see Remark~\ref{rem:MidNotWeak}.

  \item[\circled{\scriptsize{7}}] Proposition~\ref{pr:middle}, 3)
    \item[\circled{\scriptsize{8}}] This is heuristic statement. See~Remark~\ref{rem:MidNotWeak}.(ii).
  \item[\circled{\scriptsize{9}}] Proposition~\ref{pr:strong}, 5)
\end{enumerate}

\section{Time consistency for stochastic processes}\label{s:tc.sp}

We preserve the same names for various types of time consistency for both the random variables and the stochastic processes. However, we stress that the nature of time consistency for stochastic processes is usually much more intricate. If $\varphi$ is an LM-measure, and $V\in\bV^p$, then in order to compare $\varphi_{t}(V)$ and $\varphi_{s}(V)$, for $s>t$, one also needs to take into account the cash flows between times $t$ and $s$.

In order to account for the intermediate cash flows, we modify appropriately the concept of the update rule.
\begin{definition}\label{def:UR.local}
The family $\mu=\{\mu_{t,s}:\, t,s\in\bT,\, t<s\}$  of maps $\mu_{t,s}:\bar{L}^{0}_{s}\times \cX\to\bar{L}^{0}_{t}$ is called a {\it generalized update rule} if for any $X\in\cX$ the family $\mu(\cdot,X)=\{\mu_{t,s}(\cdot,X):\, t,s\in\bT,\, t<s\}$ is an update rule.
\end{definition}
Note that the update rule introduced  in Definition~\ref{def:tmbdotc} may be considered as the generalized update rule, which  is constant with respect to $X$, i.e., $\mu(\cdot,X)=\mu(\cdot,Y)$ for any $X,Y\in\cX$. In what follows, if there is no ambiguity, we drop the term {\it generalized}.

As before, we say that the update rule $\mu$ is {\it $s$-invariant}, if there exists a family $\{\mu_{t}\}_{t\in\bT}$ of maps $\mu_{t}:\bar{L}^{0}\times\cX \to\bar{L}^{0}_{t}$, such that $\mu_{t,s}(m_s,X)=\mu_{t}(m_s,X)$ for any $s,t\in\bT$, $s>t$, $X\in\cX$, and  $m_s\in\bar{L}^{0}_{s}$.

We now arrive at the corresponding definition of time-consistency.
\begin{definition}
Let $\mu$ be a generalized update rule. We say that the dynamic LM-measure $\varphi$ is {\it $\mu$-acceptance (resp. $\mu$-rejection) time consistent} if
\begin{equation}\label{eq:timeConPro}
\varphi_{s}(X)\geq m_{s} \quad(\textrm{resp. } \leq)\quad \Longrightarrow\quad \varphi_{t}(X)\geq \mu_{t,s}(m_{s},X)\quad(\textrm{resp. } \leq),
\end{equation}
for all $s,t\in\bT$, $s>t$, $X\in \cX$, and $m_{s}\in \bar{L}^{0}_{s}$. In particular, if property \eqref{eq:timeConPro} is satisfied for $s=t+1$, $t=0,\ldots,T$,  then we say that $\varphi$ is {\it one-step $\mu$-acceptance (resp. one-step $\mu$-rejection) time consistent}.
\end{definition}

Throughout this section, we assume that $\cX=\bV^{p}$.\footnote{We recall that the elements of $\bV^p$ are interpreted as discounted dividend processes.} We will focus our attention on one-step update rules $\mu$ such that
\begin{equation}\label{eq:prTOrv.f}
\mu_{t,t+1}(m,V)=\tilde \mu_{t,t+1}(m)+f(V_{t}), \quad t=0,\ldots,T-1,
\end{equation}
where $\tilde \mu$ is the one-step update rule for random variables, and $f:\bar{\bR}\to \bar{\bR}$ is a Borel measurable function such that $f(0)=0$.
Property \eqref{eq:prTOrv.f} is postulated primarily to allow establishing a direct connection  between our results and the existing literature.
Moreover, when using one-step update rules of form \eqref{eq:prTOrv.f}, the one-step time consistency for random variables is a particular case  of one-step time consistency for stochastic processes by considering cash flows  with only terminal payoff, namely stochastic processes such that $V=(0,\ldots,0,V_T)$.

Finally, we note that for update rules, which admit the so called nested composition property~(cf. \cite{RuszczynskiShapiro2006,Ruszczynski2010}),
\begin{equation}\label{eq:RvToPr}
{\mu_{t,s}(m,V)=\mu_{t,t+1}(\mu_{t+1,t+2}(\ldots\mu_{s-2,s-1}(\mu_{s-1,s}(m,V),V)\ldots V),V),}
\end{equation}
we have that $\mu$-acceptance (resp. $\mu$-rejection) time consistency is equivalent to one step $\mu$-acceptance (resp. $\mu$-rejection) time consistency. This is {another}  reason why we consider only one step update rules for stochastic processes.

\subsection{Weak time consistency}

We start with the following definition.

\begin{definition}\label{type.of.cons.proc.weak}
A dynamic LM-measure $\varphi$ on $\bV^p$ is {\it weakly acceptance (resp. weakly rejection) time consistent} if
\[
\varphi_{t}(V)\geq \Essinf_{t}\varphi_{t+1}(V)+V_t,\quad (\textrm{resp.}\quad \varphi_{t}(V)\leq \Esssup_{t}\varphi_{t+1}(V)+V_t\,)
\]
for any $V\in\bV^p$ and $t\in\bT$, such that $t<T$.
\end{definition}

The next result is the counterpart of Proposition~\ref{pr:weak} and Proposition~\ref{pr:weak2}.
\begin{proposition}\label{pr:proc.weak}
Let $\varphi$ be a dynamic LM-measure on $\bV^p$. The following properties are equivalent:
\begin{enumerate}[1)]
\item $\varphi$ is weakly acceptance time consistent.
\item $\varphi$ is  $\mu$-acceptance time consistent, where $\mu$ is an s-invariant update rule, given by
\[
\mu_{t}(m,V)=\Essinf_{t}m+V_t.
\]
\item For any $V\in\bV^p$ and $t<T$
\begin{equation}\label{eq:weak.radon}
\varphi_{t}(V)\geq \essinf_{Q\in\cM_t(P)}E_{Q}[\varphi_{t+1}(V)|\mathcal{F}_{t}]+V_t.
\end{equation}
\item For any $V\in \bV^p$, $t<T$, and $m_{t}\in \bar{L}^{0}_{t}$,
\[
\varphi_{t+1}(V)\geq m_{t} \Rightarrow \varphi_{t}(V)\geq m_{t}+V_t.
\]
\end{enumerate} Additionally, if $\varphi$ is a dynamic monetary risk measure, then the above properties are equivalent to
\begin{enumerate}[1)]
\item[5)] For any $V\in \bV^p$ and $t<T$,
\[
\varphi_{t+1}(V)\geq 0 \Rightarrow \varphi_{t}(V)\geq V_t.
\]
\end{enumerate}
Analogous equivalences are true for weak rejection time consistency.
\end{proposition}

The proof of Proposition~\ref{pr:proc.weak} is analogous to the proofs of Proposition~\ref{pr:weak} and Proposition~\ref{pr:weak2}, and we omit it.

As mentioned earlier, the update rule, and consequently time consistency for stochastic processes, depends also on the value of the process (the dividend paid)  at time $t$. In the case of weak time consistency this feature is interpreted as follows:
if tomorrow, at time $t+1$, we accept $V\in\bV^p$ at the level greater than $m_{t+1}\in\cF_{t+1}$, then today at time $t$, we will accept $V$ at least at the level $\Essinf_t m_{t+1}$ (i.e., the worst level of $m_{t+1}$ adapted to the information $\cF_t$) plus the dividend $V_t$ received today.

Finally, we present the counterpart of Proposition~\ref{pr:UDMprop} for the case of stochastic processes.
\begin{proposition}\label{pr:UDMprop.pr}
 Let $\phi$ be a projective update rule for random variables and let the update rule $\mu$ for stochastic processes be given by \begin{equation}\label{any.weak.pr}
\mu_{t,t+1}(m,V)=\phi_{t}(m)+V_{t},\quad m\in\bar{L}^{0}_{t+1},\ V\in\bV^p.
\end{equation}
If $\varphi$  is a dynamic one-step LM-measure on $\bV^p$, which is
$\mu$-acceptance (resp. $\mu$-rejection) time consistent, then $\varphi$ is weakly acceptance (resp. weakly rejection) time consistent.
\end{proposition}
\noindent Proposition~\ref{pr:UDMprop.pr} can be proved in a way analogous to the proof of Proposition~\ref{pr:UDMprop}.

\begin{remark}
The statement of Proposition~\ref{pr:UDMprop.pr} remains true if we replace \eqref{any.weak.pr} with
\[
\mu_{t,t+1}(m,V)=\phi_{t}(m+V_{t}),\quad m\in\bar{L}^{0}_{t+1},\ V\in\bV^p.
\]
Indeed, it is enough to note that, for any $V\in\bV^p$ and $t<T$,
\begin{align*}
\varphi_{t}(V) &\geq \mu_{t,t+1}(\varphi_{t+1}(V),V)=\phi_{t}(\varphi_{t+1}(V)+V_{t})\\
& \geq \phi_{t}(\Essinf_t[\varphi_{t+1}(V)+V_{t}])=\Essinf_t[\varphi_{t+1}(V)+V_{t}]\geq \Essinf_t\varphi_{t+1}(V)+V_{t}.
\end{align*}
\end{remark}

\subsection{Semi-weak time consistency}
In this section, we introduce the concept of semi-weak time consistency for stochastic processes. We have not discussed semi-weak time consistency in the case of random variables, since, in that case,  semi-weak time consistency coincides with the weak time consistency.

As it was shown, \cite{BCZ2010}, none of  the forms of time consistency existing in the literature at the time when that paper was written  were suitable for scale-invariant maps such as acceptability indices. In fact, even the weak acceptance and the weak rejection time consistency for stochastic processes (as defined in the present paper)  are too strong in the case of scale invariant maps.
This is a reason why we introduce yet a weaker notion of time consistency, which we will refer to as \textit{semi-weak acceptance} and \textit{semi-weak rejection} time consistency.
The notion of semi-weak time consistency for stochastic processes, introduced next, is well suited for scale-invariant maps; we refer the reader to \cite{BCZ2010} for a detailed discussion on time consistency for such maps and their dual representations.\footnote{In \cite{BCZ2010}, the authors combine both semi-weak acceptance and rejection time consistency into one single definition and call it time consistency.}

\begin{definition}\label{def:semi}
Let $\varphi$ be a dynamic LM-measure on $\bV^p$. Then, $\varphi$ is {\it semi-weakly acceptance time consistent} if
\begin{align*}
&\varphi_{t}(V) \geq \1_{\{V_{t}\geq 0\}} \Essinf_t(\varphi_{t+1}(V))+ 1_{\{V_{t}< 0\}}(-\infty),\ \textrm{for all}\ V\in \bV^p,\ t\in\bT,\  t<T,
\end{align*}
and it is {\it semi-weakly rejection time consistent} if
\begin{align*}
\varphi_{t}(V) \leq \1_{\{V_{t}\leq 0\}}\Esssup_t(\varphi_{t+1}(V))+1_{\{V_{t}>0\}}(+\infty),\ \textrm{for all}\ V\in \bV^p,\ t\in\bT,\  t<T.
\end{align*}
\end{definition}

Clearly, weak acceptance/rejection time consistency for stochastic processes implies semi-weak acceptance/rejection time consistency.

Next, we will show that the definition of semi-weak time consistency is indeed equivalent to the time consistency introduced in  \cite{BCZ2010}.

\begin{proposition}\label{pr:proc.weak.semi}
Let $\varphi$ be a dynamic LM-measure on $\bV^p$. The following properties are equivalent
\begin{enumerate}[1)]
\item $\varphi$ is semi-weakly acceptance time consistent.
\item  $\varphi$ is one step $\mu$-acceptance time consistent, where the (generalized) update rule is given by
\[
\mu_{t,t+1}(m,V) =1_{\{V_{t}\geq 0\}}\Essinf_t m+1_{\{V_{t}< 0\}}(-\infty).
\]
\item For all $V\in \bV^p$, $t\in\bT$,  $t<T$, and $m_{t}\in \bar{L}^{0}_{t}$, such that $V_{t}\geq 0$
\[
\varphi_{t+1}(V)\geq m_{t}\quad \Longrightarrow\quad\varphi_{t}(V)\geq m_{t}.
\]
\end{enumerate}
A similar result is true for semi-weak rejection time consistency.
\end{proposition}
\noindent
For the proof, see \cite[Proposition 4.8]{BCP2014a}.

Property 3) in Proposition~\ref{pr:proc.weak.semi}, which is the definition of the (acceptance) time consistency given in \cite{BCZ2010}, best illustrates the financial meaning of semi-weak acceptance time consistency: if tomorrow we accept the dividend stream $V\in\bV^p$ at level $m_t$,  and if we get a positive dividend $V_t$ paid today at time $t$, then today we accept the cash flow $V$ at least at level $m_t$ as well. A similar interpretation is valid for semi-weak rejection time consistency.

{The next two results are important. In particular, they generalize the work done in \cite{BCZ2010} regarding duality between cash-additive risk measures and acceptability indices.}

\begin{proposition}\label{prop:DCRMtoDAI}
 Let $\{\varphi^{x}\}_{x\in\bR_{+}}$ be a decreasing family\footnote{A family, indexed by $x\in\bR_{+}$, of maps $\{\varphi_{t}^{x}\}_{t\in\bT}$,  is called {\it decreasing}, if $\varphi_{t}^{x}(X)\leq \varphi_{t}^{y}(X)$ for all $X\in\cX$, $t\in\bT$ and $x,y\in\bR_{+}$, such that $x\geq y$.} of dynamic LM-measures on $\bV^p$.
    Assume that for each $x\in\bR_{+}$, $\varphi^{x}$ is weakly acceptance (resp. weakly rejection) time consistent.
    Then, the family $\{\alpha_{t}\}_{t\in\bT}$ of maps $\alpha_{t}:\bV^p\to \bar{L}^{0}_{t}$ defined by
\begin{equation}\label{eq:DCRMtoDAI}
\alpha_{t}(V):=\esssup_{x\in \bR^{+}}\{x \1_{\{\varphi_{t}^{x}(V)\geq0\}}\},
\end{equation}
is a semi-weakly acceptance (resp. semi-weakly rejection) time consistent dynamic LM-measure.

\end{proposition}
\noindent
For the proof, see \cite[Proposition 4.9]{BCP2014a}. It will be useful to note that $\alpha_t(V)$ defined in \eqref{eq:DCRMtoDAI}
can also be written as
\begin{equation}\label{eq:DCRMtoDAI-traditional}
\alpha_{t}(V)=\sup\set {x \in \bR^{+} \mid \varphi_{t}^{x}(V)\geq 0}.
\end{equation}
\begin{proposition}\label{prop:DAItoDCRM}
Let $\{\alpha_{t}\}_{t\in\bT}$ be a dynamic LM-measure, which is independent of the past and translation invariant.\footnote{See Appendix~\ref{S:families} for details.} Assume that $\{\alpha_{t}\}_{t\in\bT}$ is semi-weakly acceptance (resp. semi-weakly rejection) time consistent.
Then, for any $x\in\bR_{+}$, the family $\varphi^{x}=\{\varphi_{t}^{x}\}_{t\in\bT}$ of maps $\varphi^{x}_{t}:\bV^p\to \bar{L}^{0}_{t}$ defined by
\begin{equation}\label{eq:DAItoDCRM}
\varphi^{x}_{t}(V):=\essinf_{c\in\bR}\{c\1_{\{\alpha_{t}(V-c1_{\{t\}})\leq x\}}\},
\end{equation}
is a weakly acceptance (resp. weakly rejection) time consistent dynamic LM-measure.
\end{proposition}
\noindent
For the proof, see \cite[Proposition 4.10]{BCP2014a}. In what follows, we will use the fact that $\varphi^{x}_{t}(V)$ defined in \eqref{eq:DAItoDCRM}
can also be written as
\begin{equation}\label{eq:DAItoDCRM-traditional}
\varphi^{x}_{t}(V)=\inf\set{c\in\bR \mid \alpha_{t}(V-c\1_{\set{t}})\leq x}.
\end{equation}

This type of dual representation, i.e.,  \eqref{eq:DCRMtoDAI} and \eqref{eq:DAItoDCRM}, or, equivalently, \eqref{eq:DCRMtoDAI-traditional} and \eqref{eq:DAItoDCRM-traditional}, first appeared in \cite{ChernyMadan2009} where the authors studied the static (one period of time) case. Subsequently, in \cite{BCZ2010}, the authors extended these results to the case of stochastic processes with special emphasis on the time consistency property.
In contrast to the results of \cite{BCZ2010}, Propositions  \ref{prop:DCRMtoDAI} and \ref{prop:DAItoDCRM} consider an arbitrary probability space, not just a finite one.

\subsection{Strong time consistency}

Let us start with the definition of strong time consistency.

\begin{definition}\label{type.of.cons.proc.strong}
Let $\varphi$ be a dynamic LM-measure on $\bV^p$. Then $\varphi$ is said to be
{\it strongly time consistent} if
\[
V_{t}=V'_{t}\ \textrm{ and } \ \varphi_{t+1}(V)= \varphi_{t+1}(V') \quad\Longrightarrow \quad\varphi_{t}(V)= \varphi_{t}(V'),
\]
for any $V,V'\in \bV^p$ and $t\in\bT$, such that $t<T$.
\end{definition}

Now, let us present the counterpart of Proposition~\ref{pr:strong}.

\begin{proposition}\label{pr:strong.pr}
Let $\varphi$ be a dynamic LM-measure on $\bV^{p}$, which is independent of the past. The following properties are equivalent:
\begin{enumerate}[1)]
\item $\varphi$ is strongly time consistent.
\item There exists an update rule $\mu$ such that: for any $t\in\bT'$, $m\in\bar{L}^{0}_{t}$, and $V,V'\in\bV^p$, satisfying  $V_{t}=V'_{t}$, we have $\mu_{t,t+1}(m,V)=\mu_{t,t+1}(m,V')$; the family $\varphi$ is both one-step $\mu$-acceptance and one-step $\mu$-rejection time consistent.
\item There exists an update rule $\mu$ such that for any $t<T$ and  $V\in\bV^p$
\[
\varphi_{t}(V)=\mu_{t,t+1}(\varphi_{t+1}(V),1_{\set{t}}V_t).
\]
\end{enumerate}
\end{proposition}

As in the case of random variables, strong time consistency is usually considered for dynamic monetary risk measures on $\bV^{\infty}$. In this case, additional equivalent properties can be established.
For brevity, we skip the details, and only show the general idea for deriving a litany of equivalent properties.
This idea is rooted in a specific construction of strongly time consistent dynamic LM-measures.

\begin{corollary}\label{cor:strong.proc2}
Let $\mu$ be a update rule for random variables. Let $\widetilde{\varphi}$ be a dynamic LM-measure on $\bV^\infty$ given by
\[
\begin{cases}
\tilde{\varphi}_{T}(V) & =V_T\\
\tilde{\varphi}_{t}(V) & =\mu_{t,t+1}(\tilde\varphi_{t+1}(V))+V_t,
\end{cases}
\]
Then, $\tilde\varphi$ is a strongly time consistent dynamic LM-measure on $\bV^\infty$.
\end{corollary}
{For a more detailed explanation of this idea and other equivalent properties see, e.g., \cite{CheriditoKupper2006} or \cite{RuszczynskiShapiro2006}.}

\subsection{Other types of time consistency}
Other types of time consistency for stochastic processes may be defined in analogy to what is done in Section~\ref{S:other.forms} for the case of random variables.
For brevity, we limit our discussion here to the update rules derived from dynamic LM-measures.

First, given a dynamic LM-measure $\varphi$ on $\bV^p$, we  denote by $\widetilde{\varphi}$ the family of maps $\widetilde{\varphi}_{t}: L_{t+1}^{p}\to \bar{L}^{0}_{t}$  given by
\begin{equation}\label{eq:tilde.varphi}
\widetilde{\varphi}_{t}(X):=\varphi_{t}(1_{\{t+1\}}X),\ \textrm{for}\ t\in\bT'.
\end{equation}
Since $\varphi$ is monotone and local on $\bV^{p}$, then, clearly,  $\widetilde{\varphi}_{t}$ is local and monotone on $L_{t+1}^{p}$.

Next, for any $t\in\bT'$, we extend  $\widetilde{\varphi}_{t}$ to $\bar{L}_{t+1}^{0}$, preserving locality and monotonicity (see Remark~\ref{rem:phiPlus.Ext}), and this extension produces a one-step update rule.

For example, the middle acceptance time consistency is obtained by taking the update rule $\mu$ given as
 \[
 \mu_{t,t+1}(m,V)=\widetilde{\varphi}^{-}_{t}(m+V_{t}),\quad t\in\bT',
 \]
 where $\widetilde{\varphi}_{t}^{-}:\bar{L}^{0}_{t+1}\to \bar{L}^{0}_{t}$ is defined as in \eqref{eq:minus}, with the sets $\cY^{-}_{A}(X)$ replaced by
$$
\cY^{-}_{t,A}(X):=\{Y\in L^{p}_{t+1} \mid \ \1_{A}Y \leq \1_{A}X\}, \quad X\in \bar{L}^{0}_{t+1}.
$$

\subsection{Taxonomy of results}\label{sec:ConcludingRemarks}

\begin{figure}
\caption{Summary of results for acceptance time consistency for stochastic processes }
\begin{tikzpicture}[scale=1.0,transform shape]\label{chart:sp}
\path node (p0) [practica1]
	{$\varphi_{t+1}(V)\geq 0 \Rightarrow \varphi_t(V)\geq V_t$ \\
	\tiny{if $\varphi$ is additionally a monetary utility measure}};
\path  (p0.south) +(0.0,-1.3) node (p1) [practica1]
	{$\varphi_{t+1}(V)\geq m_t \Rightarrow \varphi_t(V)\geq m_t+V_{t}$};
\path (p1.south)+(0.0,-1.7) node (p2) [practica1]
	{ {\scriptsize Dynamic LM-Measure $\varphi$ is }\\
	\textbf{Weakly Accept Consist }\\
	{\scriptsize if $\varphi_{t}(V)\geq \Essinf_{t}\varphi_{t+1}(V)+V_t$}};
\path (p2.south)+(0.0,-1.7) node (p3) [practica1]
	{ {\scriptsize Dynamic LM-Measure $\varphi$ is }  \\
	\textbf{Semi-weakly Accept Consist }\\
	{\scriptsize if $\varphi_{t}(V) \geq \1_{\{V_{t}\geq 0\}} \Essinf_t(\varphi_{t+1}(V))+ 1_{\{V_{t}< 0\}}(-\infty)$}};
\path (p0.east)+(4.70, 0.0) node (p8) [practica1]
	{{\scriptsize $\varphi$ is one-step $\mu$ -- accept consist and \\
	$\mu_{t,t+1}(m,V)=\phi_{t}(m)+V_{t}$ ($\phi$ is projective)}};
\path (p2.east)+(4.7, -1.6) node (p4) [practica1]
{ {\scriptsize $\varphi$ is one step $\mu$ --  accept and $\mu$ -- reject consist \\
	and $\mu_{t,t+1}(m,V)=\mu_{t,t+1}(m,1_{\set{t}}V_{t})$}};
;
\path (p2.east)+(4.7, -4.0) node (p5) [practica1]
	{{\scriptsize Dynamic LM-Measure $\varphi$ is }  \\
	\textbf{Strongly Consist }\\
	{\scriptsize if $\varphi_{t+1}(V)= \varphi_{t+1}(V')\Rightarrow \varphi_t(V)= \varphi_t(V')$,}\\
	{\scriptsize  for $V, V'\in\cX$, such that $V_{t}=V'_{t}$}};
\path (p2.east)+(4.5,+0.3) node (p9) { {\scriptsize if $\mu_{t,t+1}(m,V)=\phi_{t}(m)+V_{t}$ and $\phi$ is projective}};
\draw[<->,thick] (p0.south) --  node[anchor=west] {\rectangled{\scriptsize{1}}} (p1.north) ;
\draw[<->,thick] (p1.south) -- node[anchor=west] {\rectangled{\scriptsize{2}}} (p2.north);
\draw[<-,thick] (p2.south) -- node[anchor=west] {\rectangled{\scriptsize{3}}} (p3.north);
%
\draw[<->,thick] (p5.north) -- node[anchor=west] {\rectangled{\scriptsize{6}}} (p4.south);
%
\path[->,draw,thick] (p8.west) -- ($(p8.west)!.5!(p0.east)$) -- node[anchor=west] {\hspace{-.15cm} \rectangled{\scriptsize{4}}} ($($(p2.west)!($(p8.west)!.5!(p0.east)$)!(p2.east)$)+(0,0.3)$) --($(p2.east)+(0,0.3)$);
\path[->,draw,thick] ($(p4.north)+(-0.8,0)$) -- node[anchor=east] {\rectangled{\scriptsize{5}}} ($($(p2.west)!($(p4.north)+(-0.8,0)$)!(p2.east)$)+(0,0)$) -- ($(p2.east)+(0,0.0)$);
\end{tikzpicture}
\label{diagram.pr}
\end{figure}
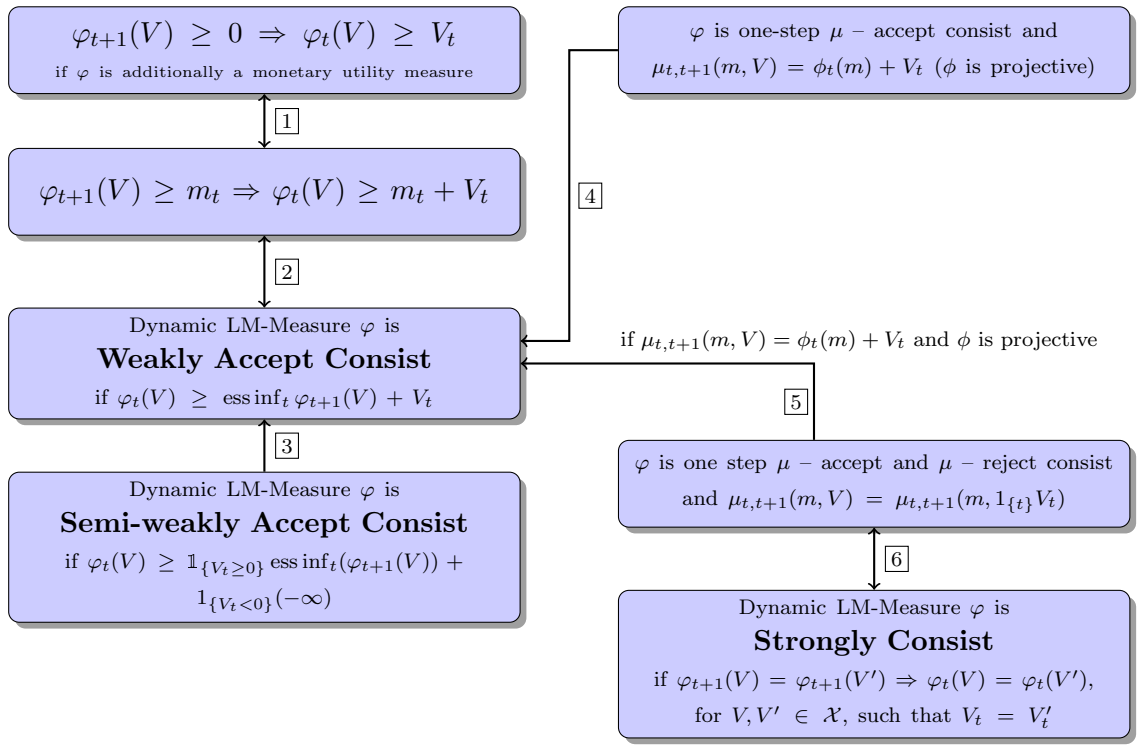

In Flowchart~\ref{chart:sp}, we summarize the results surveyed in Section \ref{s:tc.sp}.
 We label  each arrow (implication or equivalence) in the flowchart with numbers in squares and we relate the labels to the relevant results. Additionally, we provide comments on converse implications whenever appropriate.

\begin{enumerate}
  \item[\rectangled{\scriptsize{1}}] Proposition~\ref{pr:proc.weak}, 5)
  \item[\rectangled{\scriptsize{2}}] Proposition~\ref{pr:proc.weak}, 4)
  \item[\rectangled{\scriptsize{3}}] Proposition~\ref{pr:proc.weak.semi}, 3)
  \item[\rectangled{\scriptsize{4}}] Proposition~\ref{pr:UDMprop.pr}
  \item[\rectangled{\scriptsize{5}}] Proposition~\ref{pr:UDMprop.pr}, and see also \rectangled{\scriptsize{4}}.
  \item[\rectangled{\scriptsize{6}}] Proposition~\ref{pr:strong.pr}.
\end{enumerate}
\begin{remark}
The converse of implications \rectangled{\scriptsize{4}} and \rectangled{\scriptsize{5}} in Flowchart~2 do not hold true in general;
one can use the same counterexamples as in the case of random variables.
For a counterexample showing that the converse of \rectangled{\scriptsize{3}} does not hold true in general, see Example~\ref{ex:2}.
\end{remark}

\section{Examples}\label{sec:Examples}
In this section, we present examples that illustrate the different types of time consistency for dynamic risk measures and dynamic performance measures, as well as the relationships between them.

We recall that according to the convention adopted in this paper, the dynamic LM-measures representing risk measures are the negatives of their classical counterparts.  With this understanding, in the titles of the examples representing risk measures below we will skip the term ``negative''.

\begin{example}[Value at Risk ($\var$)]\label{ex:var} Let $\cX=L^{0}$ and $\alpha\in (0,1)$. We denote by $\varphi^{\alpha}_{t}(X)$ an $\cF_{t}$-conditional  $\alpha$-quantile of $X$
\begin{equation}\label{eq:var.dyn}
\varphi^{\alpha}_{t}(X):= \esssup\set{ Y\in L_{t}^0 \mid P[X\leq Y \mid  \cF_{t}]\leq\alpha }.
\end{equation}
According to our convention, the conditional $\var$ is defined by $\var^\alpha_t(X):= -\varphi^\alpha_t(X)$.

 The family of maps $\{\varphi^{\alpha}_{t}\}_{t\in\bT}$ is a dynamic monetary utility measure. It is well known that $\{\varphi^{\alpha}_{t}\}_{t\in\bT}$ is not strongly time consistent; see \cite{CheriditoStadje2009} for details. However, it is both weakly acceptance and weakly rejection time consistent. Indeed, if  $\varphi^{\alpha}_{s}(X)\geq 0$, for some $s>t$, and $X\in L^0$, then for any $\epsilon<0$ we get $P[X\leq\epsilon\mid\cF_s] = E[\1_{\set{X\leq \epsilon}}|\cF_{s}]\leq\alpha$, and
\[
E[\1_{\set{X\leq \epsilon}}|\cF_{t}]=E[E[\1_{\set{X\leq \epsilon}}|\cF_{s}]\,|\,\cF_t]\leq\alpha.
\]
Since $\epsilon<0$ was chosen arbitrarily, we get $\varphi^{\alpha}_{t}(X)\geq 0$, and thus, in view of  Proposition~\ref{pr:weak2}, $\{\varphi^{\alpha}_{t}\}_{t\in\bT}$ is  weakly acceptance time consistent.

Now, let us assume that  $\varphi^{\alpha}_{s}(X)\leq 0$.
Then, due to the locality of the conditional expectation, we have that $E[\1_{\set{X\leq \epsilon}}|\cF_{s}]>\alpha$, for any $\epsilon>0$.
In fact, if $P[E[\1_{\set{X\leq \epsilon}}|\cF_{s}]>\alpha]<1$, then there exists an $\cF_s$-measurable set $A$ with positive measure on which
\[
E[\1_{\set{X\leq \epsilon}}|\cF_{s}]\leq \alpha.
\]
Taking any $Y'\in L_s^0$ such that $E[\1_{\set{X\leq Y'}}|\cF_{s}]\leq \alpha$,  we know that for $\cF_s$-measurable random variable $Z:=\1_A\epsilon+\1_{A^c} Y'$ we get
\[
E[\1_{\set{X\leq Z}}|\cF_{s}]=\1_AE[\1_{\set{X\leq Z}}|\cF_{s}]+\1_{A^c}E[\1_{\set{X\leq Z}}|\cF_{s}]=\1_AE[\1_{\set{X\leq \epsilon}}|\cF_{s}]+\1_{A^c}E[\1_{\set{X\leq Y'}}|\cF_{s}]\leq\alpha.
\]
Thus,
\[
0\geq \esssup\set{ Y\in L_{s}^0 \mid P[X\leq Y \mid  \cF_{s}]\leq\alpha } \geq Z,
\]
which leads to the contradiction.

Consequently, for any $Y\in L^0_t$ and $\epsilon>0$,  we get
\[
E[\1_{\set{X\leq Y}}|\cF_{t}]\geq E[\1_{\set{X\leq \epsilon<Y}}|\cF_{t}]=E[\1_{\set{X\leq \epsilon}}\1_{\set{Y>\epsilon}}|\cF_{t}]=\1_{\set{Y>\epsilon}}E[E[\1_{\set{X\leq \epsilon}}|\cF_{s}]|\cF_{t}],
\]
and, consequently,  $E[\1_{\set{X\leq Y}}|\cF_{t}]>\alpha$ on $\cF_t$-measurable set $\{Y>\epsilon\}$.
Hence,
\[
\varphi^{\alpha}_{t}(X)=\esssup\set{ Y\in L_{t}^0 \mid E[\1_{\set{X\leq Y}}|\cF_{t}]\leq\alpha }\leq \esssup\set{ Y\in L_{t}^0 \mid Y\leq \epsilon }=\epsilon.
\]
Since $\epsilon>0$ was chosen arbitrary, we conclude that $\varphi^{\alpha}_{t}(X)\leq 0$, thus $\{\varphi^{\alpha}_{t}\}_{t\in\bT}$ is weakly rejection time consistent.

\end{example}

\begin{example}[Conditional Weighted Value at Risk]\label{ex:1}
Let $\cX=L^{0}$.
For a fixed $\alpha\in (0,1)$, we consider the family of sets $\{D^{\alpha}_{t}\}_{t\in\bT}$ defined by
\begin{equation}\label{eq:det.sets.coh}
D^{\alpha}_{t}:=\{Z\in L^{1}: 0\leq Z\leq\alpha^{-1},\ E[Z|\cF_{t}]=1\},\end{equation}
and we set
\begin{equation}\label{eq:robus2t}
\varphi^{\alpha}_{t}(X):=\essinf_{Z\in D^{\alpha}_{t}}E[ZX|\cF_{t}],\quad\quad t\in\bT,\ X\in L^{0}.
\end{equation}
The family of maps $\{\varphi^{\alpha}_{t}\}_{t\in\bT}$ is a dynamic coherent  utility measure (see, e.g.,~\cite{ChernyWVAR2006} for details).
Moreover, it is submartingale time consistent.
Indeed, let $t,s\in\bT$, $s>t$. Clearly,  $D^{\alpha}_{s}\subseteq D^{\alpha}_{t}$, thus
\begin{equation}\label{eq:ex1.1}
\varphi^{\alpha}_{t}(X)=\essinf_{Z\in D^{\alpha}_{t}}E[ZX|\mathcal{F}_{t}]\leq \essinf_{Z\in D^{\alpha}_{s}}E[ZX|\mathcal{F}_{t}]
= \essinf_{Z\in D^{\alpha}_{s}}E[E[ZX|\cF_{s}]|\mathcal{F}_{t}].
\end{equation}
Now, using the fact that $D^{\alpha}_{s}$ is $L^{1}$-closed (see~\cite{ChernyWVAR2006} for details), for any $X\in L^{0}$, there exist $Z_{X}^{*}\in D^{\alpha}_{s}$ such that $\varphi^{\alpha}_{s}(X)=E[Z_{X}^{*}X|\cF_{s}]$. This implies that
\begin{equation}\label{eq:ex1.2}
\essinf_{Z\in D_{s}^\alpha}E[E[ZX|\cF_{s}]|\mathcal{F}_{t}]\leq E[E[Z_{X}^{*}X|\cF_{s}]|\mathcal{F}_{t}]=E[\essinf_{Z\in D_{s}^\alpha}E[ZX|\cF_{s}]|\cF_{t}]=E[\varphi^{\alpha}_{s}(X)|\cF_{t}].
\end{equation}
Combining \eqref{eq:ex1.1} and \eqref{eq:ex1.2}, we conclude that $\varphi^\alpha$ is  submartingale time-consistent.
In particular, by Remark~\ref{rem:subThenWeak}, $\varphi^\alpha$ is  also weakly  rejection time consistent.

On the other hand, as shown in~\cite{ArtznerDelbaenEberHeathKu2007}, $\varphi^{\alpha}$ is neither middle  rejection time consistent nor weakly  acceptance time consistent.
\end{example}

\begin{example}[Dynamic $\tvar$ Acceptability Index for Processes]\label{ex:2}
Tail Value at Risk Acceptability Index was introduced in \cite{ChernyMadan2009}, as an example of static scale invariant performance measure for the case of random variables.
Here, along the lines of  \cite{BCZ2010},  we extend this notion to the dynamic setup and apply it to the case of stochastic processes.
Let $\cX=\bV^{0}$, and for a fixed $\alpha\in (0,1]$, we consider the sets $\{D^{\alpha}_{t}\}_{t\in\bT}$ defined in \eqref{eq:det.sets.coh}.
We consider the distortion function $g(x)=\frac{1}{1+x}$, $x\in\bR^{+}$, and we define $\rho^{x}=\{\rho_{t}^{x}\}_{t\in\bT}, \ x\in\bR_+$, as follows
\begin{equation}
\rho^{x}_{t}(V)=\essinf_{Z\in D^{g(x)}_{t}}E[Z\sum_{i=t}^{T}V_{i}|\cF_{t}], \quad V\in\bX, \ t\in\cT.
\end{equation}
Then, $\rho^x$ is an increasing (with respect to $x$) family of dynamic coherent  utility measures for processes, and the map $\alpha=\{\alpha_{t}\}_{t\in\bT}$ given by
\begin{equation}\label{eq:DAIT}
\alpha_{t}(V)=\sup\{x\in\bR_{+} \mid  \rho_{t}^{x}(V)\leq0\},
\end{equation}
is a dynamic acceptability index for processes (see~\cite{ChernyMadan2009} and \cite{BCZ2010}). Moreover,
\begin{equation}\label{eq:rhox}
  \rho_t^x(V) = \inf\set{c\in\bR \mid \alpha_t(V+c\1_{\set{t}})\geq x}.
\end{equation}
Clearly,  \eqref{eq:DAIT} and \eqref{eq:rhox}

are the counterparts of \eqref{eq:DCRMtoDAI-traditional} and \eqref{eq:DAItoDCRM-traditional}, respectively.

Considering the above, then, similarly to Example~\ref{ex:1}, one can show that $\rho^{x}$ is weakly   rejection time consistent, but it is not weakly acceptance time consistent, for any fixed $x\in\bR_{+}$,
and hence, by Proposition~\ref{prop:DCRMtoDAI} and Proposition~\ref{prop:DAItoDCRM},  $\alpha$ is semi-weakly  rejection time consistent but not semi-weakly  acceptance time consistent.

\end{example}

\begin{example}[Dynamic RAROC] \label{ex:3}
The Risk Adjusted Return On Capital (RAROC) is a popular scale invariant measure of performance; see \cite{ChernyMadan2009} for static RAROC and \cite{BCZ2010} for its extension to the dynamic setup.
We consider the space $\cX=\bV^{1}$, and for a fixed  $\alpha\in (0,1)$ the dynamic RAROC is defined as follows
\begin{equation}\label{eq:dRAROC}
\varphi_{t}(V):=\left\{
\begin{array}{ll}
\frac{E[\sum_{i=t}^{T}V_{i}|\cF_{t}]}{-\rho^{\alpha}_{t}(V)}&\quad \textrm{if }  E[\sum_{i=t}^{T}V_{i}|\cF_{t}]>0,\\
0 &\quad \textrm{otherwise},
\end{array}\right.
\end{equation}
when $\rho_t^\alpha(V)<0$, where $\rho^{\alpha}_{t}(V)=\essinf\limits_{Z\in D^{\alpha}_{t}}E[Z\sum_{i=t}^{T}V_{i}|\cF_{t}]$, and
$\{D^{\alpha}_{t}\}_{t\in\bT}$ given in \eqref{eq:det.sets.coh}, and $\varphi_{t}(V)=+\infty$, if $\rho_{t}(V)\geq 0$.
It was shown in \cite{BCZ2010} that $\varphi$ is a dynamic acceptability index for processes.
Moreover, for any fixed $t\in\bT$, we have that (cf. \cite{BCP2013})
$$
\varphi_{t}(V)=\sup\{x\in\bR_{+}: \phi^{x}_{t}(V)\geq 0 \},
$$
where $\phi^{x}_{t}(V)=\essinf\limits_{Z\in \cB^{x}_{t}}E[Z(\sum_{i=t}^{T}V_{i})|\cF_{t}]$, with
$\cB_{t}^{x}=\{Z\in L^{1}: Z=\frac{1}{1+x}+\frac{x}{1+x}Z_{1},\ \textrm{for some } Z_{1}\in D_{t}^{\alpha}\}$.
It is easy to check that the family $\{\varphi^{x}_{t}\}_{t\in\bT}$ is a dynamic coherent  utility measure for processes, and by similar arguments as in Example~\ref{ex:1}, we get that for any fixed $x\in\bR_{+}$,  $\varphi_{t}^{x}$ is weakly  rejection time consistent, but not weakly   acceptance time consistent.
Since $1\in D_{t}^{\alpha}$, it follows that $\{\phi_{t}^{x}\}_{t\in\bT}$ is increasing in $x\in\bR_{+}$, and by similar arguments as in Example~\ref{ex:2}, we conclude that $\varphi$ is semi-weakly rejection time consistent, but not semi-weakly  acceptance time consistent.
\end{example}

\begin{example}[Dynamic Gain Loss Ratio] \label{ex:4}
Dynamic Gain Loss Ratio (dGLR) is another popular measure of performance, which essentially improves on some drawbacks of Sharpe Ratio (such as penalizing for positive returns), and it is given by the ratio of expected return over expected losses. Formally, for $\cX=\bV^{1}$, dGLR is defined as
\begin{equation}\label{e:dGLT}
\varphi_{t}(V):=\left\{
\begin{array}{ll}
\frac{E[\sum_{i=t}^{T}V_{i}|\cF_{t}]}{E[(\sum_{i=t}^{T}V_{i})^{-}|\cF_{t}]}, &\quad \textrm{if }  E[\sum_{i=t}^{T}V_{i}|\cF_{t}]>0,\\
0, &\quad \textrm{otherwise}.
\end{array}\right.
\end{equation}
For various properties and dual representations of dGLR see \cite{BCZ2010,BCDK2013}.
In \cite{BCZ2010},  assuming that $\Omega$ is finite, the authors showed that dGLR is both semi-weakly acceptance and semi-weakly rejection time consistent.
For the sake of completeness, we will show here that dGLR is semi-weakly acceptance time consistent.

Assume that $t\in\bT'$,  and $V\in\cX$. In view of Definition~\ref{def:semi}, it is enough to show that
\begin{equation}\label{eq:ex4.1}
\varphi_{t}(V)\geq \1_{\{V_{t}\geq 0\}}\Essinf_{t}(\varphi_{t+1}(V))+\1_{\{V_{t}< 0\}}(-\infty).
\end{equation}
On the set ${\{V_{t}< 0\}}$, the inequality \eqref{eq:ex4.1} is trivial.
Since $\varphi_{t}$ is non-negative and local, without loss of generality, we may assume that $\Essinf_{t}(\varphi_{t+1}(V))>0$.
Since, $\varphi_{t+1}(V)\geq\Essinf_{t}(\varphi_{t+1}(V))$, we have that
\begin{equation}\label{eq:ex4.2}
E[\sum_{i=t+1}^{T}V_{i}|\cF_{t+1}]\geq \Essinf_{t}(\varphi_{t+1}(V))\cdot E[(\sum_{i=t+1}^{T}V_{i})^{-}|\cF_{t+1}].
\end{equation}
Using \eqref{eq:ex4.2} we obtain
\begin{align}
\1_{\{V_{t}\geq 0\}}E[\sum_{i=t}^{T}V_{i}|\cF_{t}] & \geq \1_{\{V_{t}\geq 0\}}E[E[\sum_{i=t+1}^{T}V_{i}|\cF_{t+1}]|\cF_{t}]\nonumber\\
& \geq \1_{\{V_{t}\geq 0\}}\Essinf_{t}(\varphi_{t+1}(V))\cdot E[\1_{\{V_{t}\geq 0\}}E[(\sum_{i=t+1}^{T}V_{i})^{-}|\cF_{t+1}]|\cF_{t}]\nonumber\\
& \geq \1_{\{V_{t}\geq 0\}}\Essinf_{t}(\varphi_{t+1}(V))\cdot E[(\sum_{i=t}^{T}V_{i})^{-}|\cF_{t}].\label{eq:ex4.3}
\end{align}
{Note that $\Essinf_{t}(\varphi_{t+1}(V))>0$ implies that $\varphi_{t+1}(V)>0$, thus $E[\sum_{i=t+1}^{T}V_{i}|\cF_{t+1}]>0$.
Hence, on the set $\set{V_t\geq0}$, we have
\[
E[ \sum_{i=t}^T V_i |\cF_t ] \geq  E[E[\sum_{i=t+1}^{T}V_{i}|\cF_{t+1}]|\cF_t]>0.
\]
We conclude the proof by combining the last inequality with \eqref{eq:ex4.3}. }
\end{example}

\begin{example}[Dynamic Entropic Risk Measure] \label{ex:5}
Entropic Risk Measure is a classical convex risk measure. The dynamic version of it (up to the negative sign) is defined as follows
\begin{equation}\label{eq:entrDRM}
\varphi^{\gamma}_{t}(X)=\left\{
\begin{array}{ll}
\frac{1}{\gamma}\ln E[\exp(\gamma X)|\mathcal{F}_{t}] &\quad \textrm{if } \gamma\neq 0,\\
E[X|\mathcal{F}_{t}] & \quad \textrm{if } \gamma=0,
\end{array}\right.
\end{equation}
where $X\in\cX=L^{\infty}$, $t\in\bT$. The parameter $\theta=-\gamma$ is commonly known as the risk-aversion parameter.
It can be proved that for $\gamma\leq 0$, the map $\varphi^{\gamma}_{t}$  is a dynamic  concave utility measure, and that for any $\gamma\in\bR$, the map $\varphi^{\gamma}$ is strongly time consistent (cf. \cite{KupperSchachermayer2009}).  Since it is also cash-additive, strong time consistency implies both weak rejection and weak acceptance time consistency.
Moreover (see \cite{KupperSchachermayer2009,BCP2013} for details), $\{\varphi^{\gamma}_{t}\}_{t\in\bT}$  is supermartingale time consistent if and only if $\gamma\geq 0$, and submartingale time consistent if and only if  $\gamma\leq 0$.
\end{example}

\begin{example}[Dynamic Entropic Risk Measure with non-constant risk aversion] \label{ex:6}
One can generalize the Dynamic Entropic Risk Measure \eqref{eq:entrDRM} by taking time dependent risk aversion parameters. Let
\begin{equation}
\varphi^{\gamma_{t}}_{t}(X)=\left\{
\begin{array}{ll}
\frac{1}{\gamma_{t}}\ln E[\exp(\gamma_{t} X)|\mathcal{F}_{t}] &\quad \textrm{if } \gamma_{t}\neq 0,\\
E[X|\mathcal{F}_{t}] & \quad \textrm{if } \gamma_{t}=0,
\end{array}\right.
\end{equation}
where $\{\gamma_{t}\}_{t\in\bT}$ is such that $\gamma_{t}\in L^{\infty}_{t}$, $t\in\bT$.
It has been shown in \cite{AcciaioPenner2010} that $\{\varphi_{t}^{\gamma_{t}}\}_{t\in\bT}$ is strongly time consistent if and only if $\{\gamma_{t}\}_{t\in\bT}$ is a constant process, and that it is middle acceptance time consistent if and only if $\{\gamma_{t}\}_{t\in\bT}$  is a non-increasing process, and that it is middle rejection time consistent if and only if $\{\gamma_{t}\}_{t\in\bT}$  is non-decreasing.
\end{example}

\begin{example}[Dynamic Certainty Equivalent] \label{ex:7}
Dynamic Certainty Equivalents form a large class of dynamic risk measures, with  Dynamic Entropic Risk Measure being a particular case. In this example, following \cite{KupperSchachermayer2009}, we consider an infinite time horizon, and take $\bT=\bN$ and $\cX=L^{\infty}$.
We let $U:\bar{\mathbb{R}}\rightarrow\bar{\mathbb{R}}$ be a strictly increasing and continuous function on  $\bar{\mathbb{R}}$, i.e., strictly increasing and continuous on $\bR$, with $U(\pm\infty)=\lim_{n\to\pm\infty}U(n)$. Let $\varphi=\{\varphi_{t}\}_{t\in\bT}$ be defined by
\begin{equation}\label{eq:DCE}
\varphi_{t}(X)=U^{-1}(E[U(X)|\mathcal{F}_{t}]),\quad\quad X\in \cX,\ t\in\mathbb{T}.
\end{equation}
It is easy to check that $\varphi$ is a strongly time consistent dynamic LM-measure. It belongs to the class of so called dynamic certainty equivalents \cite{KupperSchachermayer2009}. In \cite{KupperSchachermayer2009}, the authors showed that every dynamic LM-measure, which is finite, normalized, strictly monotone, continuous, law invariant, admits The Fatou property, and is strongly time consistent, can be represented as \eqref{eq:DCE} for some $U$. We also refer to \cite{BiaginiBion-Nadal2012} for a more general approach to dynamic certainty equivalents (e.g., by using stochastic utility functions $U$),  and to \cite{BCDK2013} for the  definition of certainty equivalents for processes.
\end{example}

\begin{example}[Dynamic Risk Sensitive Criterion]\label{ex:DRSC}
In \cite{BCP2013} the authors introduced the notion of the Dynamic Limit Growth Index (dLGI) that is designed to measure the long-term performance of a financial portfolio in discrete time. The dynamic analog of Risk Sensitive Criterion (cf. \cite{Whittle1990,Bielecki2003,DavisLleo2014} and references therein) is a particular case of dLGI.
We consider an infinite time horizon setup, $\bT=\bN$, and the following space  suitable for our needs
$\bV_{\ln}^{p}:=\{(W_{t})_{t\in\bT}: W_{t}>0, \ln W_{t}\in L^{p}_{t}\}$.
To be consistent with \cite{BCP2013}, we view the elements of $\cX$ as cumulative value processes of  portfolios of some financial securities, which have integrable growth expressed as cumulative log-return (note that everywhere else in the present paper, the stochastic processes represent dividend streams). Let $\varphi^{\gamma}=\{\varphi^{\gamma}_{t}\}_{t\in\bT}$ be defined by
\begin{equation}\label{e:DRSC}
\varphi^{\gamma}_{t}(W)=\left\{
\begin{array}{ll}
\liminf_{T\rightarrow\infty }\frac{1}{T}\frac{1}{\gamma}\ln E[W_{T}^{\gamma}|\mathcal{F}_{t}],&\quad \textrm{if $\gamma\neq 0$},\\
\liminf_{T\rightarrow\infty }\frac{1}{T} E[\ln W_{T}|\mathcal{F}_{t}], &\quad \textrm{if $\gamma=0$},
\end{array}\right.
\end{equation}
where $\gamma$ is a fixed real number.
It was proved in \cite{BCP2013}  that $\varphi^{\gamma}$  is a dynamic measure of performance, and it is $\mu$-acceptance time consistent with respect to $\mu_{t}(m)=E[m|\cF_{t}], \ t\in\bT$, if and only if $\gamma>0$, and $\mu$-rejection time consistent, with respect to $\mu$ if and  only if $\gamma<0$.
\end{example}

\subsection{Taxonomy of examples}

The following table is meant to help the reader to navigate through the examples presented above relative to various types of time consistency studied in this paper.
We will use the following abbreviations for time consistency:
WA - weak acceptance;
WR - weak rejection;
sWA - semi-weak acceptance;
sWR - semi-weak rejection;
MA - middle acceptance;
MR - middle rejection;
STR - strong;
Sub - submartinagle;
Sup - supermartinagle.

If a cell is marked with the check mark, that means that the corresponding property of time consistency is satisfied; otherwise the property is not satisfied in general.

We note that Example~\ref{ex:DRSC} is not represented in the table due to the distinct nature of the example. The dGLI evaluates a process $V$, but it does it through a limiting procedure, which really amounts to evaluating the process through its ``values at $T=\infty$.'' We refer the reader to \cite{BCP2013} for a detailed discussion on various properties of this measure.

\bigskip

\begin{tabular}{|cc|c|c|c|c|c|c|c|c|c|c|}
  \hline
                          &  &  $\cX$  &  WA                   & WR & sWA & sWR & MA & MR & STR & Sub & Sup   \\ \hline \hline
  Example~\ref{ex:var} & & $L^p$ & $\checkmark$   &  $\checkmark$ & $\checkmark$ & $\checkmark$  &   &   &   &    &    \\ \hline
  Example~\ref{ex:1} & & $L^p$ &  &  $\checkmark$ &  & $\checkmark$  &   &   &   &   $\checkmark$ &    \\ \hline
  Example~\ref{ex:2} & & $\bV^p$ &              &   &  & $\checkmark$  &   &   &   &   &   \\ \hline
  Example~\ref{ex:3} & & $\bV^p$ &             &   & & $\checkmark$  &   &   &   &   &   \\ \hline
  Example~\ref{ex:4} & & $\bV^p$ &              &   & $\checkmark$ &  $\checkmark$ &   &   &   &   &   \\ \hline

  \multirow{2}{*}{Example~\ref{ex:5}} & \multicolumn{1}{c|}{$\gamma\geq 0$}&  \multirow{2}{*}{$L^p$} & $\checkmark$ &  $\checkmark$& $\checkmark$ &   $\checkmark$  & $\checkmark$ & $\checkmark$ &  $\checkmark$&   & $\checkmark$ \\

                                 & \multicolumn{1}{c|}{$\gamma\leq 0$}& & $\checkmark$ & $\checkmark$  & $\checkmark$ & $\checkmark$  & $\checkmark$  & $\checkmark$  & $\checkmark$  & $\checkmark$  &    \\ \hline
  \multirow{2}{*}{Example~\ref{ex:6}} & \multicolumn{1}{c|}{$\gamma_{t} \downarrow$}&  \multirow{2}{*}{ $L^p$ } & $\checkmark$ &   & $\checkmark$  &      & $\checkmark$  &   &    &   & $\checkmark^{*}$ \\
                                 & \multicolumn{1}{c|}{$\gamma_{t} \uparrow$}& &   & $\checkmark$ &   & $\checkmark$ &    & $\checkmark$ &    & $\checkmark^{**}$ &    \\ \hline
  Example~\ref{ex:7} & &  $L^p$ & $\checkmark$& $\checkmark$ & $\checkmark$ & $\checkmark$ & $\checkmark$ & $\checkmark$ & $\checkmark$ &  &  \\ \hline
                                \multicolumn{12}{r}{{\footnotesize $^{*}$if $\gamma_{t}\geq 0$,\ $^{**}$if $\gamma_{t}\leq 0$ }}
\end{tabular}

\begin{appendix}
\section{Appendix}\label{S:A}
Here we provide a brief exposition of the three fundamental concepts used in the paper: the dynamic LM-measures, the conditional essential suprema/infima, and the LM-extensions.
\subsection{Dynamic LM-measures}\label{S:families}
Let $\cX$ denote the space of random variables or adapted stochastic processes as described in Section~\ref{sec:prelims}. 

We start with listing additional properties that may be enjoyed by a dynamic LM-measure.
Let $\varphi$ be a dynamic LM-measure. We say that $\varphi$ is
\begin{itemize}
\item {\it Super-additive} if $\varphi_t(X+Y)\geq \varphi_t(X)+\varphi_t(Y)$;
\item{\it Normalized} if $\varphi_t(0)=0$;
\item {\it Cash-additive} if $\varphi(X+m1_{\{t\}})=\varphi_t(X)+m$;
\item {\it Quasi-concave} if $\varphi_t(\lambda\cdot_{t}X+(1-\lambda)\cdot_{t}Y)\geq \varphi_t(X) \wedge \varphi_t(Y)$;
\item {\it Concave} if $\varphi_t(\lambda\cdot_{t}X+(1-\lambda)\cdot_{t}Y)\geq \lambda \varphi_t(X)+ (1-\lambda)\varphi_t(Y)$;
\item {\it Scale invariant} if $\varphi_t(\beta\cdot_{t}X)=\varphi_t(X)$;
\item  {\it Positively homogeneous} if $\varphi_t(\beta\cdot_{t}X)=\beta \varphi_t(X)$;
\item{\it Lower semi-continuous with respect to the topology $\eta$}, if $\{Z\in\bar{L}_{t}^{0} \mid \varphi_t(X)\leq Z\}$ is $\eta$-closed;\footnote{That is closed with respect to topology $\eta$; if $\eta$ will be clear from the context, we will simply write that $f$ is {\it lower semi-continuous}. If $\cX= L^{p}$, then we use the topology induced by $\|\cdot\|_{p}$ norm (see \cite[Appendix~A.7]{FollmerSchiedBook2004} for details).}
\item {\it Upper semi-continuous with respect to the topology $\eta$}, if $\{Z\in\bar{L}_{t}^{0} \mid \varphi_t(X)\geq Z\}$ is $\eta$-closed,
\end{itemize}
for any $X,Y\in\cX$, $t,s\in\bT$, such that $s>t$, and $m,\lambda,\beta\in L_t^{p}$, such that $0\leq\lambda\leq1$, $\beta>0$ and $\|\beta\|_{\infty}<\infty$. Moreover, if $\cX=\bV^p$, then we say that $\varphi$ is
\begin{itemize}
\item {\it Independent of the past} if  $\varphi_t(X)=\varphi_t(X-0\cdot_{t}X)$;
\item {\it Translation invariant}  if $\varphi_t(X+m1_{\set{t}})=\varphi_t(X+m1_{\set{s}})$.
\end{itemize}
These last two properties are automatically satisfied for $\cX=L^{p}$.

Most of the above properties have a natural financial interpretation. For example, {\it quasi-concavity}, {\it concavity}, or {\it super-additivity} correspond to the positive effect of portfolio diversification. See \cite{FollmerSchied2010,ChernyMadan2009} for more details and for a financial interpretation of other properties listed above.

Next, we recall the {\it Fatou property}, the {\it Lebesgue property}, as well as the {\it law-invariance property}. For simplicity, we present them only for the case of random variables.
We say that a dynamic LM-measure $\varphi$ admits
\begin{itemize}
\item {\it Fatou property}, if $\varphi_t(X)\geq \limsup_{n\to\infty}\varphi_t(X_{n})$;
\item {\it Lebesgue property}$\, $, if $\varphi_t(X)=\lim_{n\to\infty}\varphi(X_{n})$;
\item {\it Law-invariant property} if $\varphi_t(X)=\varphi_t(Y)$, whenever $Law(X)=Law(Y)$;
\end{itemize}
for any $t\in\bT$, $X,Y\in\cX$ and any dominated sequence$\, $\footnote{This means that there exist $Y\in \cX$ such that for all $n\in\bN$ we have $|X_{n}|\leq |Y|$.} $\set{X_{n}}_{n\in\bN}$ such that $X_{n}\in L^p$ and $X_{n}\xrightarrow{a.s.}X$.

\subsubsection{Classes of dynamic LM-measures}\label{S:FamUM}

We say that a dynamic LM-measure $\varphi$ is a
\begin{itemize}
\item {\it Dynamic monetary utility measure}, or just dynamic utility measure for short, if $\varphi$ is translation invariant, independent of the past, normalized, monotone, and cash-additive;

\item {\it Dynamic concave utility measure}, if $\varphi$ is a dynamic utility measure and concave;
\item {\it Dynamic coherent utility measure}, if $\varphi$ is a dynamic utility measure, is positive homogeneous, and super-additive;
\item {\it Dynamic performance measure}, if $\varphi$ is adapted, translation invariant, independent of the past, monotone increasing, and scale invariant;
\item {\it Dynamic acceptability index}, if $\varphi$ is a dynamic performance measure, and it is quasi-concave.
\end{itemize}

It needs to be stressed that in the literature, typically, the negative of the dynamic (monetary, concave, or coherent) utility measure is used and referred to as  \textit{dynamic (monetary, convex, or coherent) risk measure}.

\subsubsection{Robust representations for dynamic monetary utility measures}\label{A:robust}

Robust representations have been studied for general dynamic LM-measures, not only for dynamic monetary utility measures. However, in this paper we only use robust representations for dynamic monetary utility measures for random variables, and that is why our discussion here is limited to this case.
Consequently, we take $\cX=L^p$ for a fixed $p\in \{0,1,\infty\}$.

Let $\varphi$ be a dynamic monetary utility measure.
We associate with $\varphi$ the following families of objects:
\begin{itemize}
\item {\it acceptance  and rejection sets} denoted by $\cA=\{\cA_t\}_{t\in\bT}$ and $\cR=\{\cR_t\}_{t\in\bT}$, respectively, where
\begin{align*}
\cA_t &:=\{X\in L^p \colon \varphi_{t}(X)\geq 0\},\\
\cR_t &:=\{X\in L^p \colon \varphi_{t}(X)\leq 0\}.
\end{align*}
\item {\it conditional acceptance and conditional rejection sets} denoted by $\{\cA_{t,s}: t,s\in\bT,\, s>t\}$ and $\{\cR_{t,s}: t,s\in\bT,\, s>t\}$, respectively,  where
\begin{align*}
\cA_{t,s} &:=\{X\in L^p \cap \bar{L}^0_s \colon \varphi_{t}(X)\geq 0\},\\
\cR_{t,s} &:=\{X\in L^p \cap \bar{L}^0_s \colon \varphi_{t}(X)\leq 0\}.
\end{align*}
\item {\it minimal penalty functions} denoted by $\alpha^{\min}=\{\alpha^{\min}_t\}_{t\in\bT}$, where $\alpha^{\min}_t\colon \cM(P)\to \bar{\bR}$ is given by
\[
 \alpha^{\min}_{t}(Q):=-\essinf_{X\in \cA_{t}} E_{Q}[X\,|\,\cF_{t}].
\]
\item {\it conditional minimal penalty functions} denoted by $\{\alpha^{\min}_{t,s}: t,s\in\bT,\, s>t\}$, where $\alpha^{\min}_{t,s}\colon \cM(P)\to \bar{L}^{0}_{t}$ is given by
\[
 \alpha^{\min}_{t,s}(Q):=-\essinf_{X\in \cA_{t,s}} E_{Q}[X\,|\,\cF_{t}].
\]
\end{itemize}

\noindent The following important definition is frequently used in this paper.

\begin{definition}
Let $\varphi$ be a dynamic monetary utility measure. We call $\varphi$ {\it representable}, if
\begin{equation}\label{eq:representable}
\varphi_t(X)=\essinf_{Q\in\cM(P)}\big(E_{Q}[X\mid \cF_{t}]+\alpha_{t}^{\min}(Q)\big),
\end{equation}
for any $X\in \cX$.
\end{definition}
This type of representation is called robust or numerical representations.
Moreover, such representation characterizes dynamic concave utility measures that admit the Fatou property.

\subsection[Conditional expectation and conditional essential supremum/infimum]{Conditional expectation and conditional essential supremum/infimum}\label{A:cond}
 We present here some relevant properties of the generalized conditional expectation and conditional essential superemum and infimum, in the context of $\bar{L}^{0}$.
\begin{proposition}\label{pr:condexp}
For any $X,Y\in\bar{L}^{0}$ and $s,t\in\bT$, $s>t$, it holds that
\begin{enumerate}[1)]
\item $E[\lambda X|\cF_{t}]\leq \lambda E[X|\cF_{t}]$ for $\lambda\in L^{0}_{t}$, and $E[\lambda X|\cF_{t}]=\lambda E[X|\cF_{t}]$ for $\lambda\in L^{0}_{t}$, $\lambda\geq 0$;
\item $E[X|\cF_{t}]\leq E[E[X|\cF_{s}]|\cF_{t}]$, and $E[X|\cF_{t}]=E[E[X|\cF_{s}]|\cF_{t}]$ for $X\geq 0$;
\item $E[X|\cF_{t}]+E[Y|\cF_{t}]\leq E[X+Y|\cF_{t}]$, and $E[X|\cF_{t}]+E[Y|\cF_{t}]=E[X+Y|\cF_{t}]$, if $X,Y\geq 0$.
\end{enumerate}
\end{proposition}
For the proof, see \cite[Proposition A.1]{BCP2014a}.

\begin{remark}
All inequalities in Proposition~\ref{pr:condexp} can be strict.
Assume that $t=0$ and $k,s\in\bT$, $k>s > 0$, and let $\xi \in L^{0}_{k}$ be such that $\xi=\pm1$, $\xi$ is independent of $\cF_s$, and $P(\xi=1)=P(\xi=-1)=1/2$.
We consider $Z\in L_{s}^{0}$ such that $Z\geq 0$, and $E[Z]=\infty$.
By taking $\lambda=-1$, $X=\xi Z$ and $Y=-X$, we get strict inequalities in 1), 2), and 3).
\end{remark}

We proceed with presenting a  definition of the conditional essential infimum and supremum that is equivalent to the one presented in Section~\ref{sec:prelims}, cf. \eqref{eq:essinf.b}. We start by recalling the definition of the conditional essential infimum for bounded random variables. For $X\in L^{\infty}$ and $t\in\bT$, we denote by $\Essinf_{t}X$ the unique (up to a set of measure zero), the $\cF_{t}$-measurable random variable, such that for any $A\in\cF_{t}$, the following equality holds true
\begin{equation}\label{eq:essinf.b}
\essinf_{\omega\in A}X=\essinf_{\omega\in A}(\Essinf_{t}X).
\end{equation}
We  call this random variable the \textit{$\cF_{t}$-conditional essential infimum of $X$}. Accordingly, we define $\Esssup_{t}(X):=-\Essinf_{t}(-X)$, the  \textit{$\cF_{t}$-conditional essential supremum of $X\in L^\infty$}. The reader is referred to~\cite{BarronCardaliaguetJensen2003} for a proof of the existence and uniqueness of the conditional essential supremum/infimum.

Consequently, for any $t\in\bT$ and $X\in\bar{L}^{0}$, we define the $\cF_{t}$-conditional essential infimum by
\begin{equation}\label{eq:A22}
\Essinf_{t}X:=\lim_{n\to\infty}\Big[\Essinf_{t}(X^{+}\wedge n)\Big]-\lim_{n\to\infty}\Big[\Esssup_{t}(X^{-}\wedge n)\Big].
\end{equation}
Respectively, we put $\Esssup_{t}(X):=-\Essinf_{t}(-X)$.

\begin{proposition}\label{pr:essinf}
For any $X,Y\in \bar{L}^{0}$, $s,t \in \bT, s\geq t$, and $A\in\cF_{t}$ the following properties hold,
\begin{enumerate}[1)]
\item $\essinf_{\omega\in A}X=\essinf_{\omega\in A}(\Essinf_{t}X)$;
\item If $\essinf_{\omega\in A}X=\essinf_{\omega\in A}U$ for some $U\in \bar{L}^{0}_t$, then $U=\Essinf_{t}X$;
\item $X\geq \Essinf_{t}X$;
\item  If $Z\in\bar{L}^{0}_{t}$, is such that $X\geq Z$, then $\Essinf_{t}X\geq Z$;
\item If $X\geq Y$, then $\Essinf_{t}X\geq \Essinf_{t}Y$;
\item $\1_{A}\Essinf_{t}X=\1_{A}\Essinf_{t}(\1_{A}X)$;
\item $\essinf_{s}X\geq \Essinf_{t}X$;
\end{enumerate}
Analogous results are true for $\{\Esssup_{t}\}_{t\in\bT}$.
\end{proposition}
The proof for the case $X,Y\in L^{\infty}$ can be found in~\cite{BarronCardaliaguetJensen2003}. Since for any $n\in\bN$ and $X,Y\in \bar{L}^{0}$, we get $X^{+}\wedge n\in L^{\infty}$, $X^{-}\wedge n\in L^{\infty}$, and $X^{+}\wedge X^{-}=0$, the extension of the proof to the case $X,Y\in \bar{L}^{0}$ is straightforward.

 It is worth mentioning that properties 3) and 4) from Proposition~\ref{pr:essinf} imply that the conditional essential infimum $\essinf_{t}(X)$  can be defined as the largest $\cF_{t}$-measurable random variable, which is smaller than $X$ (cf.~\cite{BarronCardaliaguetJensen2003}).

Next, we define the generalized versions of $\essinf$ and $\esssup$ of a (possibly uncountable) family of random variables.
For $\{X_{i}\}_{i\in I}$, where $X_{i}\in\bar{L}^{0}$, we let
\begin{equation}\label{eq:cond222}
\essinf_{i\in I}X_{i}:=\lim_{n\to\infty}\Big[\Essinf_{i\in I}(X_{i}^{+}\wedge n)\Big]-\lim_{n\to\infty}\Big[\Esssup_{i\in I}(X_{i}^{-}\wedge n)\Big].
\end{equation}
Note that, in view of \cite[Appendix~A]{KaratzasShreve1998},  $\Essinf_{i\in I}X_{i}\wedge n $ and $\Esssup_{i\in I}X_{i}\wedge n $ are well defined, so that  $\essinf_{i\in I}X_{i}$ is well defined. It needs to be observed that the operations of the right-hand side of \eqref{eq:cond222} preserve measurability. In particular, if $X_{i}\in\cF_{t}$ for all $i\in I$, then $\essinf_{i\in I}X_{i}\in\cF_{t}$.

Furthermore, if for any $i,j\in I$, there exists $k\in I$, such that $X_{k}\leq X_{i}\wedge X_{j}$, then there exists a sequence $i_n\in I, n\in\bN$, such that $\{X_{i_n}\}_{n\in\bN}$ is non-increasing and $\essinf_{i\in I}X_{i}=\inf_{n\in\bN}X_{i_n}=\lim_{n\to\infty}X_{i_n}$.
Analogous results hold true for $\esssup_{i\in I}X_{i}$.

\subsection{LM-extensions}\label{S:extensions}
In this part of the appendix, we introduce the concept of an LM-extension of a dynamic LM-measure for random variables.

\begin{definition}\label{def:ext}
Let $\varphi$ be a dynamic LM-measure on $L^p$. We call a family $\widehat{\varphi}=\{\widehat{\varphi}_{t}\}_{t\in\bT}$ of maps $\widehat{\varphi}_{t}:\bar{L}^{0}\to \bar{L}^{0}_{t}$ {\it an LM-extension of  $\varphi$}, if for any $t\in\bT$, $\widehat{\varphi}_{t}|_{\cX}\equiv\varphi_{t}$, and $\widehat{\varphi}_{t}$ is local and monotone on $\bar{L}_0$.\footnote{That is, it satisfies monotonicity and locality on $\bar{L}_{0}$, as in 5) and 6) in Proposition \ref{pr:essinf}.}
\end{definition}

We will show below that such extension exists, for which we will make use of the following auxiliary sets:
\[
\cY^{+}_{A}(X):=\{Y\in\cX\mid \ \1_{A}Y \geq \1_{A}X\},\quad\quad \cY^{-}_{A}(X):=\{Y\in\cX\mid \ \1_{A}Y \leq \1_{A}X\},
\]
defined for any $X\in\bar{L}^{0}$ and $A\in\cF$.
\begin{definition}
Let $\varphi$ be a dynamic LM-measure. The collection of functions $\varphi^{+}=\{\varphi_{t}^{+}\}_{t\in\bT}$, where $\varphi^{+}_{t}:\bar{L}^{0}\to \bar{L}^{0}_{t}$ is defined as\footnote{We will use the convention $\esssup{\emptyset}=-\infty$ and $\essinf{\emptyset}=\infty$.}
\begin{equation}\label{eq:plus}
\varphi^{+}_{t}(X):=\essinf_{A\in\cF_{t}}\Big[\1_{A}\essinf_{Y\in \cY^{+}_{A}(X)}\varphi_{t}(Y)+\1_{A^{c}}(+\infty)\Big],
\end{equation}
is called the {\it upper LM-extension of $\varphi$}. Respectively,  the collection of functions $\varphi^{-}=\{\varphi_{t}^{-}\}_{t\in\bT}$, where $\varphi^{-}_{t}:\bar{L}^{0}\to \bar{L}^{0}_{t}$, and
\begin{equation}\label{eq:minus}
\varphi^{-}_{t}(X):=\esssup_{A\in\cF_{t}}\Big[\1_{A}\esssup_{Y\in \cY^{-}_{A}(X)}\varphi_{t}(Y)+\1_{A^{c}}(-\infty)\Big],
\end{equation}
is called the {\it lower LM-extension of $\varphi$}.
\end{definition}

The next result shows that $\varphi^{\pm}$ are two ``extreme'' extensions, and any other extension is sandwiched between them.

\begin{proposition}\label{pr:upper.ext}
Let $\varphi$ be a dynamic LM-measure. Then, $\varphi^{-}$ and $\varphi^{+}$ are LM-extensions of $\varphi$. Moreover, let $\widehat{\varphi}$ be an LM-extension of $\varphi$. Then, for any $X\in\bar{L}^{0}$ and $t\in\bT$,
\begin{equation}\label{eq:bounds}
\varphi_{t}^{-}(X)\leq\widehat{\varphi}_{t}(X)\leq \varphi_{t}^{+}(X).
\end{equation}
\end{proposition}

Clearly, in general,  the maps \eqref{eq:plus} and \eqref{eq:minus} are not equal, and thus the extensions of an LM-measure are not unique.

\begin{remark}\label{rem:phiPlus.Ext}
Let $t\in\bT$ and $\cB\subseteq\bar{L}^{0}$ be such that, for any $A\in\cF_{t}$, $\1_{A}\cB\subseteq\cB$, and $\1_{A}\cB+\1_{A^{c}}\cB\subseteq\cB$.
As a generalization of  Proposition~\ref{pr:upper.ext}, one can show that for any $\cF_{t}$-local and monotone mapping\footnote{That is, $\cF_{t}$-local and monotone on $\cB$.} $f:\cB\to \bar{L}_{t}^{0}$, the maps $f^\pm$ defined analogously  as in ~\eqref{eq:plus}~and~\eqref{eq:minus} are extensions of  $f$ to $\bar{L}^{0}$,  preserving locality and monotonicity.
\end{remark}

\begin{remark}\label{rem:robustExt}
For a large class of LM-measures, as mentioned earlier, there exists a ``robust representation'' type theorem---essentially a representation, via convex duality, as a function of conditional expectation.  We refer the reader to \cite{BCDK2013} and references therein, where the authors present a general robust representation for
dynamic quasi-concave upper semi-continuous LM-measures. Hence, an alternative construction of extensions can be obtained through the robust representations of LM-measures, by considering conditional expectations defined on the extended real number line, etc.
\end{remark}

\section{Proofs}\label{A:proofs}

\subsection*{Proof of Proposition~\ref{pr:strong}.}\label{pr:strong.a}
\begin{proof}

\noindent $1)\Rightarrow 2)$. Let $t,s\in\bT$ be such that $s>t$, and consider the following set
$$
\cX_{\varphi_{s}}=\{X\in \bar{L}^{0}\mid X=\varphi_{s}(Y)\ \textrm{for some}\ Y\in\cX\},
$$
where $\cX=L^p$. From 1), for any $X,Y\in\cX$, such that  $\varphi_{s}(X)=\varphi_{s}(Y)$, we get $\varphi_{t}(X)=\varphi_{t}(Y)$.
Next, we define the map $\phi_{t,s}: \cX_{\varphi_{s}}\to \bar{L}^{0}_{t}$ as follows: for any $X'\in\cX_{\varphi_s}$
\begin{equation}\label{eq:strong:proof1}
\phi_{t,s}(X')=\varphi_{t}(X),\quad X\in\cX,
\end{equation}
where $X\in\cX$ is such that $X'=\varphi_{s}(X)$. In view of the definition of $\cX_{\varphi_{s}}$ and strong time consistency of $\varphi$, the map $\phi_{t,s}$ is well-defined.

 Since there exists $Z\in\cX$, such that $\varphi_{s}(Z)=0$ (see  property \eqref{eq:normalization}), using locality of $\varphi$, we get that for any $X\in\cX_{\varphi_{s}}, \ A\in\cF_t$, there exists $Y\in\cX$, so that
\[
\1_A X = \1_A\varphi_s(Y) = \1_A\varphi_s(\1_AY) + \1_{A^c}\varphi_s(\1_{A^c} Z) = \varphi_s(\1_A Y + \1_{A^c}Z).
\]
Thus,  $\1_{A}X\in\cX_{\varphi_{s}}$, for any $A\in\cF_{t}, \ X\in\cX_{\varphi_{s}}$.
Hence, from 1) and the locality of $\varphi$, for any $X,Y\in \cX_{\varphi_{s}}$, $A\in \cF_{t}$, we get
\begin{itemize}
\item[(A)] $X\geq Y \Rightarrow \phi_{t,s}(X)\geq \phi_{t,s}(Y)$;
\item[(B)] $\1_{A}\phi_{t,s}(X)=\1_{A}\phi_{t,s}(\1_{A}X)$.
\end{itemize}
In other words, $\phi_{t,s}$ is local and monotone on $\cX_{\varphi_{s}}\subseteq \bar{L}^{0}_{s}$.
By  Remark~\ref{rem:phiPlus.Ext}), there exists an extension of  $\phi_{t,s}$, say $\widehat{\phi}_{t,s}: \bar{L}^{0}_{s}\to \bar{L}^{0}_{t}$, which is local and monotone on $\bar{L}^{0}_{s}$.
Finally, we take $\mu_{t,s}:\bar{L}^{0}_{s} \to \bar{L}^{0}_{t}$ defined by
$$
\mu_{t,s}(m):=\widehat{\phi}_{t,s}(m),\quad\quad m\in \bar{L}^{0}_{s}.
$$
Clearly, the family $\mu=\{\mu_{t,s}:\, t,s\in\bT,\, s>t\}$ is an update rule, and using \eqref{eq:strong:proof1}, we get that $\varphi$ is both $\mu$-acceptance and $\mu$-rejection time consistent.

\smallskip
\noindent $2)\Rightarrow 3)$. Let $s,t\in\bT$ and $X,Y\in\cX$ be such that $s>t$ and $\varphi_{s}(X)\geq\varphi_{s}(Y)$. From 2),\eqref{eq:accepTimeConsAlt}, and by the monotonicity of $\mu$, we have $ \varphi_{t}(X)=\mu_{t,s}(\varphi_{s}(X))\geq \mu_{t,s}(\varphi_{s}(Y))=\varphi_{t}(Y)$.
\smallskip

\noindent $3)\Rightarrow 1)$,  $4)\Leftrightarrow 2)$,  and $4)\Rightarrow 5)$ are obvious.

\smallskip
\noindent $5)\Rightarrow 4)$. Let a family $\tilde{\mu}=\{\tilde{\mu}_{t,s}:\, t,s\in\bT,\, t<s\}$ of maps $\tilde{\mu}_{t,s}:\bar{L}^{0}_{s}\to\bar{L}^{0}_{t}$ be given by
\[
\begin{cases}
\tilde{\mu}_{t,s}(\cdot) :=\mu_{t,t+1}(\cdot) & \textrm{if } s=t+1,\\
\tilde{\mu}_{t,s}(\cdot) :=\mu_{t,t+1}\circ \ldots \circ {\mu}_{s-1,s}(\cdot) & \textrm{if } s>t+1,
\end{cases}
\]
where $\mu$ is an update rule from 5). It is straightforward to check that $\tilde{\mu}$ is an update rule, and that $\varphi$ is both $\tilde{\mu}$-acceptance and $\tilde{\mu}$-rejection time consistent, which proves that 4) holds.

The proof is complete.

\end{proof}

\subsection*{Proof of Proposition~\ref{pr:coh.upd}.}\label{pr:coh.upd.a}
\begin{proof}
Let us consider $\{\phi_{t}\}_{t\in\mathbb{T}}$ as given in~(\ref{eq:robust}).

\smallskip
\noindent 1) The proof of monotonicity and locality is similar to the one for the conditional essential infimum and supremum, Proposition~\ref{pr:essinf}.
Finally, for any $t\in\bT$, $Z\in D_{t}$, and $m\in\bar{L}^{0}_{t}$, since $E[Z|\cF_{t}]=1$, we immediately get
\begin{align*}
E[Zm|\cF_{t}] = 1_{\{m\geq 0\}}m E[Z|\cF_{t}] +1_{\{m< 0\}}(-m) E[-Z|\cF_{t}]=m,
\end{align*}
and thus, $\phi_t(m)=m$, for any $m\in\bar{L}^{0}_{t}$. Hence, $\{\phi_{t}\}_{t\in\bT}$ is projective.

\smallskip
\noindent 2) Let $\varphi$ be a dynamic LM-measure which is $\phi$-rejection time consistent, and $g:\bar{\bR}\to\bar{\bR}$ be an increasing, concave function.
Then, for any $X\in\cX$, we get
\begin{equation}\label{eq:pr19.1}
g(\varphi_{t}(X))\geq  g(\phi_{t}(\varphi_{s}(X))=g(\essinf_{Z\in D_{t}}E[Z\varphi_{s}(X)|\cF_{t}])=\essinf_{Z\in D_{t}}g(E[Z\varphi_{s}(X)|\cF_{t}].
\end{equation}
 Recall that any $Z\in D_{t}$ is a Radon-Nikodym derivative of some measure $Q$ with respect to $P$, and thus we have $E[ZX|\cF_{t}]=E_{Q}[X|\cF_{t}]$.
 Hence, by Jensen's inequality,  we deduce
\begin{equation}\label{eq:pr19.2}
\essinf_{Z\in D_{t}}g(E[Z\varphi_{t}(X)|\cF_{t}])\geq \essinf_{Z\in D_{t}}E[Zg(\varphi_{t}(X))|\cF_{t}]=\phi_{t}(g(\varphi_{s}(X))).
\end{equation}
Combining \eqref{eq:pr19.1} and \eqref{eq:pr19.2}, $\phi$-acceptance time consistency of $\{g\circ\varphi_{t}\}_{t\in\bT}$ follows.
\end{proof}

\subsection*{Proof of Proposition~\ref{pr:ext}.}\label{pr:ext.a}
\begin{proof}
The first part follows immediately  from the definition of LM-extension. Clearly, projectivity of $\widehat{\varphi}$ implies that $\varphi_t(X)=X$, for $X\in\cX\cap\bar{L}^{0}_{t}$.
To prove the opposite implication, it is enough to prove that $\varphi^{+}$ and $\varphi^{-}$ are projective. Assume that $\varphi$ is such that $\varphi_{t}(X)=X$, for $t\in\bT$ and $X\in L^p\cap\bar{L}^{0}_{t}$.
Let $X\in\bar{L}^{0}_{t}$. For any $n\in\bN$, we get
$$
\1_{\{n\geq X\geq -n\}}\varphi^{+}_{t}(X)=\1_{\{n\geq X\geq -n\}}\varphi^{+}_{t}(\1_{\{n\geq X\geq -n\}}X)=\1_{\{n\geq X\geq -n\}}\varphi_{t}(\1_{\{n\geq X\geq -n\}}X)=\1_{\{n\geq X\geq -n\}}X.
$$
Thus, on set $\bigcup_{n\in\bN}\{-n\leq X\leq n\}=\{-\infty<X<\infty\}$, we have
\begin{equation}\label{eq:varphi.plus.pr}
\varphi^{+}_{t}(X)=X, \quad\quad \textrm{for } X\in\bar{L}^{0}_{t}.
\end{equation}
 Next, for any $A\in\cF_{t}$, such that  $A\subseteq \{X=\infty\}$, we get $\cY^{+}_{A}(X)=\emptyset$, which implies $\1_{\{X=\infty\}}\varphi^{+}(X)=\infty$. Finally, for any $n\in\bR$, using locality of $\varphi^{+}_{t}$ and the fact that $n\in\cX\cap\bar{L}^{0}_{t}$, we get
$$\1_{\{X=-\infty\}}\varphi_{t}^{+}(X)\leq \1_{\{X=-\infty\}}\varphi_{t}^{+}(\1_{\{X=-\infty\}}n)=\1_{\{X=-\infty\}}\varphi_{t}(n)=\1_{\{X=-\infty\}}n,$$
which implies $\1_{\{X=-\infty\}}\varphi^{+}(X)=-\infty$. Hence, \eqref{eq:varphi.plus.pr} holds true on entire space.
The proof for $\varphi^{-}$ is analogous.
\end{proof}

\subsection*{Proof of Proposition~\ref{pr:strong.pr}.}\label{pr:strong.pr.a}
\begin{proof}
Let $\varphi$ be a dynamic LM-measure, which is independent of the past.

\smallskip
\noindent $1)\Rightarrow 2)$. Let $t\in\bT'$ and consider the following set
$$
\cX_{\varphi_{t+1}}=\{X\in \bar{L}^{0}\mid X=\varphi_{t+1}(V)\ \textrm{for some}\ V\in\bV^p\}.
$$
From 1), for any $V,V'\in\cX$, such that  $\varphi_{t+1}(V)=\varphi_{t+1}(V')$ and $V_{t}=V'_{t}$,
we get $\varphi_{t}(X)=\varphi_{t}(Y)$. Thus, using the independence of the past of $\varphi$, there exists a map $\phi_{t,t+1}: \cX_{\varphi_{t+1}}\times L^{p}_{t}\to \bar{L}^{0}_{t}$ such that
$$\phi_{t,t+1}(\varphi_{t+1}(X),Y_{t})=\varphi_{t}(X-1_{\set{t}}(X_{t}-Y_{t})),\quad X\in\cX.$$
Next, since there exists $Z\in\cX$, such that $\varphi_{t+1}(Z)=0$, using the locality of $\varphi$, we get that for any $X\in\cX_{\varphi_{t+1}}, \ A\in\cF_t$, there exist $Y\in\cX$, so that
$$
\1_A X = \1_A\varphi_{t+1}(Y) = \1_A\varphi_{t+1}(\1_A\cdot_{t+1}Y) + \1_{A^c}\varphi_{t+1}(\1_{A^c}\cdot_{t+1} Z) = \varphi_{t+1}(\1_A\cdot_{t+1}Y + \1_{A^c}\cdot_{t+1}Z).
$$
Thus,  $\1_{A}X\in\cX_{\varphi_{t+1}}$, for any $A\in\cF_{t}, \ X\in\cX_{\varphi_{t+1}}$.
Hence, from 2) and the locality of $\varphi$, for any $X,X'\in \cX_{\varphi_{t+1}}$, $Y_{t}\in L^{p}_{t}$ and $A\in \cF_{t}$, we get
\begin{itemize}
\item[(A)] $X\geq X' \Rightarrow \phi_{t,t+1}(X,Y_{t})\geq \phi_{t,t+1}(X',Y_{t})$;
\item[(B)] $\1_{A}\phi_{t,t+1}(X,Y_{t})=\1_{A}\phi_{t,t+1}(\1_{A}X,Y_{t})$.
\end{itemize}
In other words, for any fixed $Y_{t}\in L^{p}_{t}$, $\phi_{t,t+1}(\cdot,Y_{t})$ is local and monotone on $\cX_{\varphi_{t+1}}\subseteq \bar{L}^{0}_{t+1}$.
In view of Remark~\ref{rem:phiPlus.Ext}, for any fixed $Y_{t}\in L^{p}_{t}$ there exists an extension (to $\bar{L}^{0}_{t+1}$) of  $\phi_{t,t+1}(\cdot,Y_{t})$, say $\widehat{\phi}_{t,t+1}(\cdot,Y_{t})$, which is local and monotone on $\bar{L}^{0}_{t+1}$.
Finally, we take $\mu_{t,t+1}:\bar{L}^{0}_{t+1}\times \cX\to \bar{L}^{0}_{t}$ defined by
$$
\mu_{t,t+1}(m,X):=\widehat{\phi}_{t,t+1}(m,X_{t}),\quad\quad X\in\cX, m\in \bar{L}^{0}_{t+1}.
$$
Clearly, the family $\mu_{t,t+1}$ is a (one step) update rule. Moreover, we get
$$
\mu_{t,t+1}(m,X)=\mu_{t,t+1}(m,X'),
$$
for $m\in \bar{L}^{0}_{t+1}$ and $X,X'\in\cX$, such that $X_{t}=X'_{t}$. Finally,  $\varphi$ is both $\mu$-acceptance and $\mu$-rejection time consistent, as
$$
\varphi_{t}(X)=\varphi_{t}(X-1_{\set{t}}(X_{t}-X_{t}))=\phi_{t,t+1}(\varphi_{t+1}(X),X_{t})=\mu_{t,t+1}(\varphi_{t+1}(X),X).
$$

\smallskip
\noindent $2)\Rightarrow 1)$. Assume that $\mu$ is an update rule, fulfilling 2), such that $\varphi$ is both $\mu$-acceptance and $\mu$-rejection time consistent.
Then, we get  $\varphi_{t}(X)=\mu_{t,t+1}(\varphi_{t+1}(X),Y)$, for any $t\in\bT'$, $X\in\cX$, and $Y\in\cX$, such that $X_{t}=Y_{t}$.
Let $t\in\bT'$ and $X,Y\in\cX$ be such that $X_{t}=Y_{t}$ and $\varphi_{t+1}(X)\geq \varphi_{t+1}(Y)$.
From the above, and by monotonicity of $\mu$, we have
$$
 \varphi_{t}(X)=\mu_{t,t+1}(\varphi_{t+1}(X),X)=\mu_{t,t+1}(\varphi_{t+1}(X),Y)\geq\mu_{t,t+1}(\varphi_{t+1}(Y),Y)=\varphi_{t}(Y).
$$

\smallskip
\noindent The proof of the equivalence between 2) and 3) is straightforward and hence omitted here.
\end{proof}

\subsection*{Proof of Proposition~\ref{pr:upper.ext}.}\label{pr:upper.ext.a}
\begin{proof}
We show the proof for $\varphi^{+}$ only; the proof  for $\varphi^{+}$  is similar.
Consider a fixed $t\in\bT$.

\smallskip
\noindent (Adaptivity) It is easy to note that for any $X\in\bar{L}^{0}$, and $A\in\cF_{t}$, we get
\begin{equation}\label{eq:phi.plus}
\Big[\1_{A}\essinf_{Y\in \cY^{+}_{A}(X)}\varphi_{t}(Y)+\1_{A^{c}}(\infty)\Big]\in \bar{L}^{0}_{t}.
\end{equation}
Indeed, for any $X\in\bar{L}^{0}$, $\essinf$ of the set of $\cF_{t}$-measurable random variables $\{\varphi_{t}(Y)\}_{Y\in \cY^{+}_{A}(X)}$ is $\cF_{t}$-measurable (see~\cite{KaratzasShreve1998}, Appendix A), which implies \eqref{eq:phi.plus} for any $A\in\cF_{t}$. Thus, $\varphi^{+}_{t}(X)\in\bar{L}^{0}_{t}$.

\smallskip
\noindent (Monotonicity) If $X\geq X'$ then for any $A\in\cF_{t}$ we get $\cY^{+}_{A}(X)\subseteq \cY^{+}_{A}(X')$, and consequently, for any $A\in\cF_{t}$,
$$
\1_{A}\essinf_{Y\in \cY^{+}_{A}(X)}\varphi_{t}(Y)\geq \1_{A}\essinf_{Y\in \cY^{+}_{A}(X')}\varphi_{t}(Y),
$$
which implies $\varphi_{t}^{+}(X)\geq\varphi_{t}^{+}(X')$.

\smallskip
\noindent {(Locality) Let $B\in\cF_{t}$ and $X\in\bar{L}^{0}$.
It is enough to consider $A\in\cF_{t}$, such that $\cY^{+}_{A}(X)\neq\emptyset$, as otherwise we get $\varphi_{t}^{+}(X)\equiv \infty$. For any such $A\in\cF_{t}$, we get
\begin{equation}\label{eq:ppp1pp1}
\1_{A\cap B}\essinf_{Y\in \cY^{+}_{A}(X)}\varphi_{t}(Y)=\1_{A\cap B}\essinf_{Y\in \cY^{+}_{A\cap B}(X)}\varphi_{t}(Y).
\end{equation}
Indeed, let us assume that $\cY^{+}_{A}(X)\neq\emptyset$.
As $\cY^{+}_{A}(X)\subseteq \cY^{+}_{A\cap B}(X)$, we have
$$
\1_{A\cap B}\essinf_{Y\in \cY^{+}_{A}(X)}\varphi_{t}(Y)\geq \1_{A\cap B}\essinf_{Y\in \cY^{+}_{A\cap B}(X)}\varphi_{t}(Y).
$$
On the other hand, for any $Y\in\cY^{+}_{A\cap B}(X)$, and any fixed $Z\in\cY^{+}_{A}(X)$ (note that $\cY^{+}_{A}(X)\neq \emptyset$), we get
$$
\1_{B}Y+\1_{B^{c}}Z\in \cY^{+}_{A}(X).
$$
Thus, using the locality of $\varphi_{t}$, we deduce
$$
\1_{A\cap B}\essinf_{Y\in \cY^{+}_{A\cap B}(X)}\varphi_{t}(Y)= \1_{A\cap B}\essinf_{Y\in \cY^{+}_{A\cap B}(X)}\1_{B}\varphi_{t}(\1_{B}Y+\1_{B^{c}}Z)\geq\1_{A\cap B}\essinf_{Y\in \cY^{+}_{A}(X)}\varphi_{t}(Y),
$$
which proves \eqref{eq:ppp1pp1}.
It is easy to see that $\cY^{+}_{A\cap B}(X)=\cY^{+}_{A\cap B}(\1_{B}X)$, and thus
\begin{equation}\label{eq:ppp1pp2}
\1_{A}\essinf_{Y\in \cY^{+}_{A\cap B}(X)}\varphi_{t}(Y)=\1_{A}\essinf_{Y\in \cY^{+}_{A\cap B}(\1_{B}X)}\varphi_{t}(Y).
\end{equation}
Combining \eqref{eq:ppp1pp1}, \eqref{eq:ppp1pp2}, and the fact that $\cY^{+}_{A}(X)\neq\emptyset$ implies $\cY^{+}_{A}(\1_{B}X)\neq\emptyset$, we continue
\begin{align*}
\1_{B}\varphi^{+}_{t}(X) &= \1_{B}\essinf_{A\in\cF_{t}}\Big[\1_{A}\essinf_{Y\in \cY^{+}_{A}(X)}\varphi_{t}(Y)+\1_{A^{c}}(\infty)\Big]\\
& = \1_{B}\essinf_{A\in\cF_{t}}\Big[\1_{A\cap B}\essinf_{Y\in \cY^{+}_{A}(X)}\varphi_{t}(Y)+\1_{A^{c}\cap B}(\infty)\Big]\\
& = \1_{B}\essinf_{A\in\cF_{t}}\Big[\1_{A\cap B}\essinf_{Y\in \cY^{+}_{A\cap B}(X)}\varphi_{t}(Y)+\1_{A^{c}\cap B}(\infty)\Big]\\
& = \1_{B}\essinf_{A\in\cF_{t}}\Big[\1_{A\cap B}\essinf_{Y\in \cY^{+}_{A\cap B}(\1_{B}X)}\varphi_{t}(Y)+\1_{A^{c}\cap B}(\infty)\Big]\\
& =\1_{B}\essinf_{A\in\cF_{t}}\Big[\1_{A}\essinf_{Y\in \cY^{+}_{A}(\1_{B}X)}\varphi_{t}(Y)+ \1_{A^{c}}(\infty)\Big]\\
& = \1_{B}\varphi_{t}^{+}(\1_{B}X).\\
\end{align*}}
\smallskip
\noindent (Extension) If $X\in\cX$, then for any $A\in\cF_{t}$, we get $X\in\cY^{+}_{A}(X)$. Thus,
$$\varphi^{+}_{t}(X)=\essinf_{A\in\cF_{t}}\Big[\1_{A}\essinf_{Y\in \cY^{+}_{A}(X)}\varphi_{t}(Y)+\1_{A^{c}}(\infty)\Big]=\essinf_{A\in\cF_{t}}\Big[\1_{A}\varphi_{t}(X)+\1_{A^{c}}(\infty)\Big]=\varphi_{t}(X).$$
As the above results are true for any $t\in\bT$, thus we have proved that $\varphi^{+}$ is an extension of $\varphi$.
Let us now show \eqref{eq:bounds} for $\varphi^{+}$.

Let $\widehat{\varphi}$ be an extension of $\varphi$, and let $X\in\bar{L}^{0}$ and $t\in\bT$.
Due to monotonicity and locality of $\widehat{\varphi}_{t}$, for any $A\in\cF_{t}$ and $Y\in\cY_{A}^{+}(X)$, we get $\1_{A}\widehat{\varphi}_{t}(X)\leq \1_{A}\widehat{\varphi}_{t}(Y)$. Thus, recalling that $\essinf\emptyset=\infty$, we have
\begin{equation}\label{eq:ext1.pp2}
\widehat{\varphi}_{t}(X)\leq \1_{A}\essinf_{Y\in\cY_{A}^{+}(X)}\widehat{\varphi}_{t}(Y)+\1_{A^{c}}(\infty)=\1_{A}\essinf_{Y\in\cY_{A}^{+}(X)}\varphi_{t}(Y)+\1_{A^{c}}(\infty).
\end{equation}
Since \eqref{eq:ext1.pp2} holds true for any $A\in\cF_{t}$, we conclude that
$$
\widehat{\varphi}_{t}(X)\leq \essinf_{A\in\cF_{t}}\Big[\1_{A}\essinf_{Y\in\cY_{A}^{+}(X)}\varphi_{t}(Y)+\1_{A^{c}}(\infty)\Big]=\varphi^{+}_{t}(X).
$$
The proof of the second inequality is analogous.
\end{proof}

\end{appendix}

\section*{Acknowledgments}
Tomasz R. Bielecki and Igor Cialenco acknowledge support from the NSF grant  DMS-1211256.
Part of the research was performed while Igor Cialenco was visiting the Institute for Pure and Applied Mathematics (IPAM), which is supported by the National Science Foundation.
Marcin Pitera acknowledges the support by Project operated within the Foundation for Polish Science IPP Programme ``Geometry and Topology in Physical Model'' co-financed by the EU European Regional Development Fund, Operational Program Innovative Economy 2007-2013. The authors would like to thank Marek Rutkowski for stimulating discussions and helpful remarks.
We would also like to thank the anonymous referees, the associate editor and the editor for their helpful
comments and suggestions which improved greatly the final manuscript.

\bibliographystyle{alpha}
{\small

\newcommand{\etalchar}[1]{$^{#1}$}

}

\end{document}